\documentclass[a4paper,fleqn]{cas-sc}

\usepackage[T1]{fontenc}
\usepackage[utf8]{inputenc}

\usepackage{amsmath,amssymb,mathtools}
\usepackage{graphicx}
\usepackage{flafter}
\usepackage[section]{placeins}

\usepackage{float}

\usepackage{hyperref}
\hypersetup{hidelinks}
\graphicspath{{figures/}}

\usepackage{setspace}
\usepackage[numbers,sort&compress]{natbib}

\ExplSyntaxOn
\cs_set:Npn \__first_footerline: {}
\keys_set:nn { stm / mktitle } { nologo }
\ExplSyntaxOff

\begin{document}

\let\WriteBookmarks\relax
\let\printorcid\relax
\shorttitle{Fractal Structures in Quantum Protocols}
\shortauthors{Portik et al.}
\title[mode=title]{Persistence of Fractal Structures in Iterated Quantum Protocols: Robustness to Preparation and Gate Errors}
\author[1,2]{Attila Portik}
\cormark[1]
\ead{portik.attila@wigner.hun-ren.hu}
\author[1]{Orsolya K\'alm\'an}
\author[1]{Tam\'as Kiss}
\affiliation[1]{organization={HUN-REN Wigner Research Centre for Physics},
 addressline={1525 P.O. Box 49}, city={Budapest}, country={Hungary}}
\affiliation[2]{organization={E\"otv\"os Lor\'and University},
 addressline={P.O.\ Box 32, H-1518}, city={Budapest}, country={Hungary}}
\cortext[1]{Corresponding author.}
\begin{abstract}
Repeatedly combining entangling gates on an ensemble of qubits with measurement and postselection on a part of them induces a nonlinear evolution of the unmeasured qubits' state, whose fractal convergence boundaries make the outcome sensitive to the initial state. We study a family of iterated protocols in which $n$ identical qubits are coupled by CNOT gates, $n-1$ of them are postselected, and a parametrized single-qubit unitary acts on the survivor. For orders $n=2$–$5$, we map the box-counting dimension of the pure-state Julia sets over the complex parameter plane and identify the symmetries that constrain it. Allowing mixed initial states, we find that moderate preparation noise preserves the main parameter-dependent trends of the boundary dimension over much of the studied range, and that the three-dimensional quasi-Julia boundary in the Bloch ball has a dimension that is approximately equal to the dimension of the pure-state Julia set plus one. We further model gate imperfections by an effective depolarizing channel and by fixed coherent interactions. Weak errors leave the estimated dimensions consistent with their ideal values even where the convergence regions visibly deform, and $ZZ$ errors are shown to be exactly conjugate to the ideal protocol at a rotated parameter. Finally, we analyze the postselection and storage cost of an implementation and propose a  diagnostic test  combining conditional fidelity with acceptance statistics, which we demonstrate on IQM Emerald and Rigetti Cepheus-1-108Q.
\end{abstract}
\maketitle

\section{Introduction}

Iterated nonlinear quantum protocols lie at the intersection of quantum information processing and nonlinear dynamics. Although quantum mechanics prescribes linear time evolution at the level of state vectors, repeatedly combining elementary quantum gates with measurement and postselection induces an effective nonlinear time evolution of a qubit state. Such protocols therefore realize genuinely nonlinear dynamical systems, capable of exhibiting sensitive dependence on initial conditions and chaos. For pure single-qubit states, the evolution can be described by rational maps of the complex plane whose Julia sets characterize the unstable dynamics~\cite{Kiss2006}. These Julia sets appear in the space of initial states as fractal boundaries between regions with different long-term behavior, revealing how small changes in the input can alter the outcome of the protocol. Their geometry thus provides information about both the underlying dynamics and its sensitivity to state preparation. Related measurement-induced nonlinear dynamics has also been studied in entanglement-purification protocols, where fractal structures and sensitive dependence on initial conditions can arise~\cite{Kiss2011,guan2013reexamination}.

The origin of this nonlinearity is simple. In each round, several identically prepared qubits are processed together, and only those runs that yield a particular measurement outcome are kept. Because the probability of obtaining this outcome itself depends on the input state, the resulting output state is a nonlinear function of the input, even though every gate and measurement involved is an ordinary quantum operation. The same principle of keeping only selected measurement outcomes underlies protocols for state purification and discrimination, where it is used to steer quantum states towards desired properties~\cite{BechmannP,KalmanKissJex2018}. Conditional selection is also central to entanglement purification, including protocols for bipartite systems of higher dimension~\cite{bennett1996purification,Alber2001,dur2007entanglement}.

Experimental implementations inevitably introduce imperfections into this evolution. Preparation errors change the initial states, while gate errors modify the transformation applied during the protocol. The sensitivity associated with fractal boundaries makes it necessary to determine whether their structure survives these imperfections. Previous studies have shown that, for particular quadratic and higher-order protocols, the fractal boundaries of the basins of attraction persist for mixed initial states above a critical purity~\cite{Malachov_2019,PORTIK2022127999}. Small coherent errors in the final single-qubit gate can also preserve the fractal structure while deforming the convergence regions and shifting the preparation-noise threshold~\cite{robustness2023}. These findings motivate examining how the geometry responds to different sources of noise across a broader family of protocols.

In this work, we investigate protocols that combine CNOT gates acting on $n$ identical input qubits, measurement and postselection of $n-1$ qubits, and a parametrized single-qubit unitary on the surviving qubit. We characterize the fractal boundaries for orders $n=2,3,4,$ and $5$, and examine how their estimated box-counting dimensions depend on the parameter of the single-qubit unitary. To account for imperfect preparation, we extend the analysis to mixed states, studying both constant-purity sections and the boundary embedded in the full Bloch ball. We then consider imperfections in the implementation by applying, at each iteration, an effective depolarizing channel and a fixed coherent interaction.

The numerical comparisons reveal that visible changes in the convergence regions can occur while the estimated fractal dimensions still agree with the ideal values. For the weak errors investigated here, the boundaries preserve their characteristic scaling over the accessible numerical resolutions. We use these observations to assess the persistence of fractal geometry and discuss the preparation, postselection, and storage requirements that constrain its experimental investigation. We also illustrate a fixed-point diagnostic on IQM Emerald and Rigetti Cepheus-1-108Q and compare the results with illustrative stochastic-noise simulations.

\section{Iterated nonlinear quantum protocols}

\subsection{Ideal map}

\begin{figure}[pos=htbp]
\singlespacing
\centering
\includegraphics{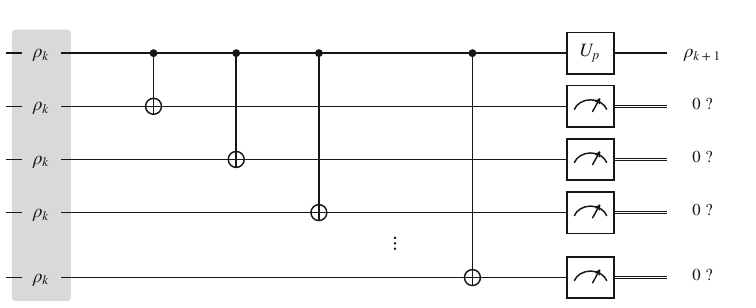}
\caption{One round of the $n$-qubit protocol. Independently prepared inputs in state $\rho_k$ undergo a CNOT network controlled by the uppermost qubit. The other $n-1$ qubits are measured, and the uppermost qubit is kept only when all outcomes are $0$. Applying $U_p$ then produces $\rho_{k+1}$.}
\label{fig:protocol_circuit}
\end{figure}

A single round of the protocol combines $n\geq2$ input qubits the first one controlling a CNOT gate on each of the remaining ones, which are then measured in the computational basis, as illustrated in Fig.~\ref{fig:protocol_circuit}. We accept only the outcome in which all $n-1$ measured qubits return $0$. A single-qubit unitary $U_p$, acting on the first (unmeasured) qubit completes the round. Repeating the procedure requires $n$ independently obtained outputs from the preceding round.

We derive the conditional evolution assuming error-free gates and measurements. The input may be pure or mixed, so the same description can subsequently be used to investigate preparation noise. For uncorrelated qubits sharing the state $\rho_k$, the joint input to round $k+1$ is
\begin{equation}\label{eq:joint_input}
\Sigma_k=\rho_k^{\otimes n}.
\end{equation}
The CNOT network and its associated unitary channel are
\begin{equation}\label{eq:ideal_core}
C_n=\prod_{j=2}^{n}\mathrm{CX}_{1\to j},
\qquad
\mathcal E_0(\Sigma)=C_n\Sigma C_n^\dagger.
\end{equation}
Selecting the required measurement outcome gives the unnormalized accepted state
\begin{equation}\label{eq:n0_def}
N_0(\rho_k)=
\operatorname{Tr}_{2,\ldots,n}
\!\left[\Pi_0\mathcal E_0(\Sigma_k)\Pi_0\right],
\qquad \text{where} \qquad
\Pi_0=I_2\otimes
\bigl(|0\rangle\langle0|\bigr)^{\otimes(n-1)}.
\end{equation}
Its trace,
\begin{equation}\label{eq:d0_def}
D_0(\rho_k)=\operatorname{Tr}N_0(\rho_k),
\end{equation}
is the probability of acceptance. The normalized state before the final gate is therefore $T_0(\rho_k)=N_0(\rho_k)/D_0(\rho_k)$.

The accepted operation has a simple interpretation in the computational basis. The measured targets return $0$ only when their original values agree with the control. Thus, the two input configurations contributing to the accepted state are $|0\rangle^{\otimes n}$ and $|1\rangle^{\otimes n}$. Keeping only these contributions gives
\begin{equation}\label{eq:ideal_accepted_output}
N_0(\rho)=
\begin{pmatrix}
\rho_{11}^{\,n}&\rho_{12}^{\,n}\\
\rho_{21}^{\,n}&\rho_{22}^{\,n}
\end{pmatrix},
\qquad
D_0(\rho)=\rho_{11}^{\,n}+\rho_{22}^{\,n},
\end{equation}
where indices $1$ and $2$ refer to $|0\rangle$ and $|1\rangle$, respectively. The resulting conditional transformation is the normalized entrywise (Hadamard) power
\begin{equation}\label{eq:T0_entrywise_general}
T_0(\rho)=
\frac{1}{\rho_{11}^{\,n}+\rho_{22}^{\,n}}
\begin{pmatrix}
\rho_{11}^{\,n}&\rho_{12}^{\,n}\\
\rho_{21}^{\,n}&\rho_{22}^{\,n}
\end{pmatrix}.
\end{equation}
For $n=2$, this reduces to the quadratic map studied in Refs.~\cite{BechmannP,PORTIK2022127999}. For physical inputs, $D_0(\rho)\geq2^{1-n}$, so normalization is always well defined.

We parametrize the final gate by a complex number $p$,
\begin{equation}\label{eq:Up}
U_p=\frac{1}{\sqrt{1+|p|^2}}
\begin{pmatrix}
1&p\\
-\bar p&1
\end{pmatrix}.
\end{equation}
At $p=1$, this gate satisfies $U_1=ZH$, where $H$ is the Hadamard gate.

Varying $p$ changes the dynamics while leaving the entangling gates and acceptance condition fixed. Including this gate, the ideal update becomes
\begin{equation}\label{eq:F0_def}
\rho_{k+1}=F_0(\rho_k)
=U_pT_0(\rho_k)U_p^\dagger.
\end{equation}
Since no assumption about purity has been made, Eq.~\eqref{eq:F0_def} applies to any input in the Bloch ball.

For pure states, it is simpler to follow the ratio of the computational-basis amplitudes. We write
\begin{equation}\label{eq:pure_state_parameter}
|\psi(z)\rangle=
\frac{|0\rangle+z|1\rangle}{\sqrt{1+|z|^2}},
\end{equation}
so that each finite $z$ corresponds to a point of the complex plane and $|1\rangle$ corresponds to $z=\infty$. For a pure input, every entry of $\rho$ is a product of two amplitudes, so the entrywise power sends $|\psi(z)\rangle$ to the pure state $|\psi(z^n)\rangle$, and the final gate produces
\begin{equation}\label{eq:pure_state_output}
U_p|\psi(z^n)\rangle=
\frac{(1+pz^n)|0\rangle+(z^n-\bar p)|1\rangle}
{\sqrt{(1+|p|^2)(1+|z|^{2n})}}.
\end{equation}
The amplitude ratio of the output therefore obeys
\begin{equation}\label{eq:fp}
z_{k+1}=f_p(z_k),
\qquad \text{with} \qquad
f_p(z)=\frac{z^n-\bar p}{1+pz^n}.
\end{equation}
A vanishing denominator represents the output $|1\rangle$. Because $z^n-\bar p$ and $1+pz^n$ have no common root for any $p$, $f_p$ is a rational map of degree $n$ on the Riemann sphere. Its iteration provides the connection between the physical protocol and the periodic orbits, unstable dynamics, and Julia sets which will be examined below~\cite{Kiss2006}.

\subsection{Sensitivity and fractal boundaries}

Iteration of $F_0$ determines which initial states approach the same asymptotic behavior. Where attracting cycles, including fixed points, exist, their basins describe the corresponding convergence regions. Initial states within a basin approach its attractor, whereas a small displacement across a basin boundary can lead to a different outcome. The arrangement of these boundaries therefore determines how accurately the input must be prepared to obtain a reproducible asymptotic result. Such sensitivity was identified in iterated postselected qubit protocols~\cite{Kiss2006}.

\begin{figure}[pos=htbp]
\singlespacing
\centering
\begin{tabular}{@{}c@{\hspace{6.4mm}}c@{\hspace{6.4mm}}c@{}}
\includegraphics{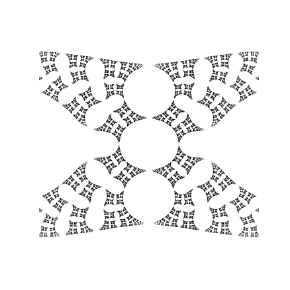} & \includegraphics{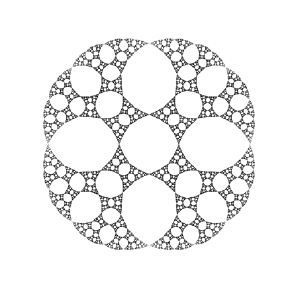} & \includegraphics{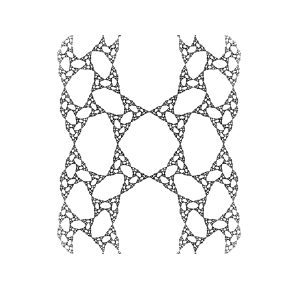} \\[1mm]
\small $n=2$, $p=0.8$. & \small $n=2$, $p=1$. & \small $n=2$, $p=1.2$. \\[1mm]
\includegraphics{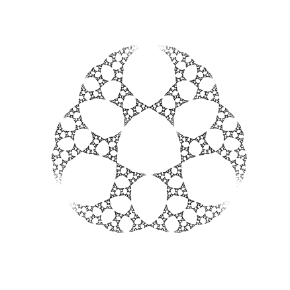} & \includegraphics{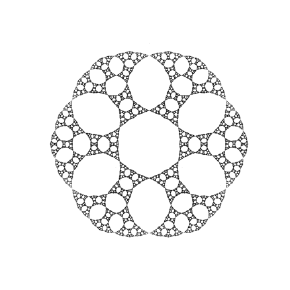} & \includegraphics{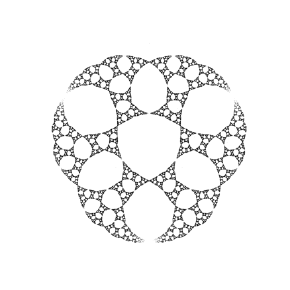} \\[1mm]
\small $n=3$, $p=0.8$. & \small $n=3$, $p=1$. & \small $n=3$, $p=1.2$.
\end{tabular}
\caption{Julia sets of $f_p$ for quadratic (upper row) and cubic (lower row) protocols. From left to right, the columns correspond to $p=0.8$, $1$, and $1.2$. Dark points belong to $J(f_p)$, and white regions belong to its complement, the Fatou set. Each panel shows the region $|\operatorname{Re} z|\leq 2$, $|\operatorname{Im} z|\leq 2$ of the complex plane.}
\label{fig:julia_examples}
\end{figure}

For pure inputs, the unstable structure is characterized by the Julia set $J(f_p)$, which is the closure of the repelling periodic points of the rational map. When attracting cycles are present, their basins have the Julia set as their boundary. This boundary is often fractal, so that different convergence regions interleave down to arbitrarily small scales. Its form depends on both the protocol order $n$ and the parameter $p$, as illustrated by the examples in Fig.~\ref{fig:julia_examples}. The Julia set is not always fractal, however. For $p=0$, the map reduces to $z\mapsto z^n$, whose Julia set is the unit circle~\cite{Milnorbook}.

Periodic convergence requires an additional distinction. Two trajectories can approach the same attracting cycle while occupying different positions along it when compared at equal iteration numbers. We therefore distinguish convergence phases as well as different cycles. This is justified because an attracting $m$-cycle of $f_p$ consists of attracting fixed points of $f_p^{m}$, and $J(f_p^{m})=J(f_p)$, so the boundaries between phase-resolved regions still lie on the Julia set. For example, the quadratic $p=1$ protocol has the attracting cycle $0\leftrightarrow-1$, which is its only attractor. Without resolving the phase, all convergent initial states would fall into a single region and no boundary would appear. Its two phase-resolved regions instead correspond to one attracting orbit, rather than two independent attractors. 

The basin description also has limits. Every attracting cycle of a rational map attracts at least one of its critical points~\cite{Milnorbook}, which for $f_p$ are $z=0$ and $z=\infty$. We refer to their orbits as critical orbits. For $p=\pm i$ and even $n$, both critical orbits land on the repelling fixed point $z=1$. The Julia set then covers the entire Riemann sphere and there are no attracting cycles, as known for the quadratic protocol~\cite{Gilyen2016,kalman2018sensitivity,Portik2026}. A method based solely on separating convergence regions therefore cannot reconstruct this case. In the following, we distinguish direct sampling of the pure-state Julia set from the classification of convergence boundaries. This becomes particularly relevant for mixed inputs, where the evolution takes place inside the Bloch ball and the geometry must be investigated beyond the pure-state surface.

\subsection{Box-counting dimension of the Julia set}
\label{sec:julia_dimension}

We quantify the parameter-dependent geometry of the Julia set through its box-counting dimension $D_{B}$. If $N_{\mathrm{box}}(\delta)$ denotes the number of boxes of side length $\delta$ required to cover the set, then
\begin{equation}\label{eq:box_dimension}
D_B=\lim_{\delta\to0}
\frac{\log N_{\mathrm{box}}(\delta)}{\log(1/\delta)},
\end{equation}
provided the limit exists. The mathematical dimension refers to the Julia set on the Riemann sphere; the finite planar grid below provides a coordinate-based estimate. Numerically, we estimate this dimension from the slope over the scales accessible to the calculation. The resulting values characterize the resolved geometry rather than the exact limiting dimension~\cite{Falconer2003}.

For each order $n=2,3,4,$ and $5$, we examine a $100\times100$ grid of parameters with $\operatorname{Re}p,\operatorname{Im}p\in[-2,2]$. For each parameter value, we sample the Julia set by backward iteration (iteration by one of the inverses of the map) from a repelling fixed point. Such a seed is selected among the roots of $pz^{n+1}-z^n+z+\bar p=0$ by requiring $|f_p'(z)|>1$.
The inverse branches are
\begin{equation}\label{eq:inverse_branches}
f_p^{-1}(y)=
e^{2\pi i m/n}
\left(\frac{y+\bar p}{1-py}\right)^{1/n},
\qquad \text{where} \qquad m=0,\ldots,n-1.
\end{equation}
At each iteration, one branch is selected with equal probability. We generate $10^7$ points and record their occupancy on a $5000\times5000$ grid. The square sampling window encloses the finite sampled points, with a $5\%$ margin added on each side. Using equal coordinate scales preserves the geometry within this window. Recording occupancy removes repeated visits but does not correct for regions that remain insufficiently sampled.

We partition the binary grid into non-overlapping square boxes and count those containing at least one sampled point. Box sizes are chosen to divide the grid into an integer number of boxes. The dimension is estimated by linear regression,
\begin{equation}\label{eq:box_fit}
\log N_{\mathrm{box}}(\delta)=c+D_B\log(1/\delta).
\end{equation}
The fit uses intermediate scales, excluding coarse boxes that poorly resolve the structure and fine boxes affected by pixel resolution or insufficient sampling. The fitting range is selected separately for each resolved structure; its extent, grid resolution, and sampling density preclude prescribing one common interval for all parameters. The plotted errors are regression standard errors. They do not include systematic changes with sampling density, covering resolution, or fitting range. These slopes provide finite-scale estimates of the box-counting dimension~\cite{Falconer2003,WU2020100106}.

Before examining the numerical diagrams, we identify the parameter transformations under which the dimension is unchanged. For $R_\omega(z)=\omega z$, where $\omega^{n-1}=1$,
\begin{equation}\label{eq:parameter_rotation}
R_\omega^{-1}\circ f_{\bar\omega p}\circ R_\omega=f_p.
\end{equation}
Parameters related by these discrete rotations generate Julia sets that differ only by a rotation of the state coordinates. Complex conjugation and the transformation $h(z)=1/z$ additionally satisfy
\begin{equation}\label{eq:parameter_reflections}
\overline{f_p(z)}=f_{\bar p}(\bar z),
\qquad
h\circ f_p=f_{-\bar p}\circ h.
\end{equation}
These relations establish reflection symmetry of the dimension diagrams about both coordinate axes. Because the associated state-space transformations are spherical isometries, they preserve the box-counting dimension. The dynamical equivalence follows from conjugacy of the maps~\cite{Milnorbook,Falconer2003}.

A further conjugacy relates the $p=1$ protocol to the protocol in which the final gate is a Hadamard gate alone. Since $U_1=ZH$, applying only $H$ induces $h_n(z)=(1-z^n)/(1+z^n)$, as studied in Refs.~\cite{Malachov_2019,PORTIK2022127999}. For even $n$, conjugation by $R(z)=-z$ gives $R^{-1}\circ h_n\circ R=f_1$, so the two dynamics are conjugate. For odd $n$, the two protocols are distinct, as their critical orbits show. The Hadamard critical orbit is $0\leftrightarrow1$, whereas the $p=1$ map has $0\to-1\to\infty\to1\to0$.

\begin{figure}[pos=htbp]
\singlespacing
\centering
\begin{tabular}{@{}c@{\hspace{6.4mm}}c@{}}
\includegraphics{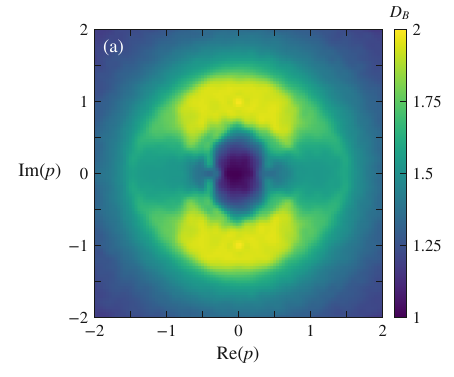} & \includegraphics{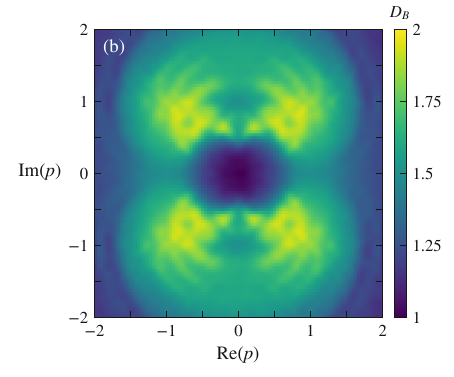} \\[1mm]
\includegraphics{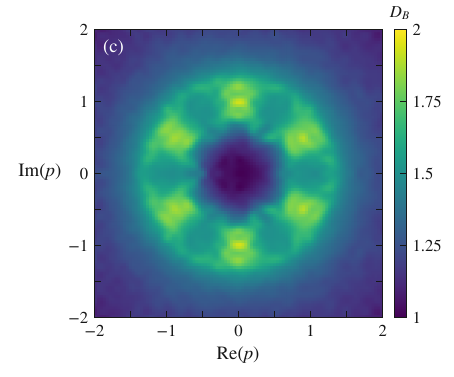} & \includegraphics{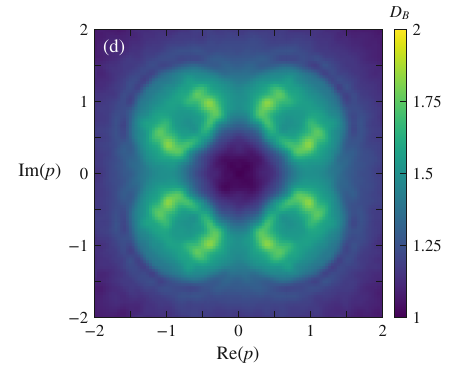}
\end{tabular}
\caption{Estimated box-counting dimension over the complex parameter plane for (a) $n=2$, (b) $n=3$, (c) $n=4$, and (d) $n=5$. Each panel covers $\operatorname{Re}p,\operatorname{Im}p\in[-2,2]$. All panels use the same color scale for the finite-resolution estimates of $D_B$.}
\label{fig:DB_full}
\end{figure}

The numerical results in Fig.~\ref{fig:DB_full} show a central region with estimates close to $D_B=1$, surrounded by regions reaching approximately $1.8$--$1.9$. At $p=0$, the Julia set is exactly the unit circle. Away from this point, both the magnitude and argument of $p$ affect the dimension, so increasing $|p|$ does not produce a universal monotonic change. The diagrams display $2,4,6,$ and $8$ prominent sectors for the four orders studied. Their relation to the exact symmetries depends on parity. Combining the parameter rotations with the reflections gives rotational order $2(n-1)$ for even $n$, but only $n-1$ for odd $n$. Thus, the observed $2(n-1)$ sectors are not always required by the rotational symmetry alone.

High-dimensional regions also require care in interpretation. For $p=\pm i$ and even $n$, the Julia set occupies the entire Riemann sphere and has dimension $2$, although there are no attracting basins to separate. These analytically known cases, discussed for the quadratic protocol in Ref.~\cite{Gilyen2016}, provide a reference for assessing the numerical dimension estimates.

Toward the outer parts of the investigated parameter domain, the estimated dimensions generally decrease. Increasing $|p|$ along a fixed direction gives, for fixed $z\neq0$,
\begin{equation}\label{eq:large_p_limit}
f_p(z)\sim-\frac{\bar p}{p}\,z^{-n}.
\end{equation}
The limiting reciprocal-power map has the unit circle as its Julia set, independently of the phase factor. This supplies a qualitative explanation for the lower dimensions observed at larger $|p|$, without establishing dimension convergence from the numerical trend alone.

\section{Effect of preparation noise}

We study preparation noise by allowing the initial single-qubit state to be mixed, while keeping the gates and measurements ideal. At each round, the participating qubits are assumed to be uncorrelated and described by the same density operator. Preparation noise therefore enters through the initial condition of the map $F_0$. This description has been used to investigate the persistence of fractal boundaries in quadratic and higher-order protocols~\cite{Malachov_2019,PORTIK2022127999}.

For mixed states, the complex amplitude ratio used for pure states is no longer sufficient. We instead introduce the Bloch vector $\mathbf r=(u,v,w)$, whose Cartesian components are the expectation values of the Pauli operators,
\begin{equation}\label{eq:bloch_coordinates}
u=\operatorname{Tr}(\rho\sigma_x),\qquad
v=\operatorname{Tr}(\rho\sigma_y),\qquad
w=\operatorname{Tr}(\rho\sigma_z).
\end{equation}
These three real coordinates determine the density operator,
\begin{equation}\label{eq:rho_bloch}
\rho=\frac12\left(I_2+u\sigma_x+v\sigma_y+w\sigma_z\right)
=\frac12
\begin{pmatrix}
1+w&u-i v\\
u+i v&1-w
\end{pmatrix}.
\end{equation}
Physical states satisfy $u^2+v^2+w^2\leq1$. Pure states lie on the unit sphere, mixed states occupy its interior (the Bloch ball), while the origin represents the maximally mixed state. The purity is given by
\begin{equation}\label{eq:purity_bloch}
P=\operatorname{Tr}(\rho^2)
=\frac{1+u^2+v^2+w^2}{2},
\end{equation}
so that a prescribed value $P>1/2$ selects a spherical surface of radius $\sqrt{2P-1}$.

Substituting this representation into $F_0$ gives the coordinate update
\begin{equation}\label{eq:bloch_update_u}
u_{k+1}=
\frac{
2\operatorname{Re}\!\left[(1-p^2)(u_k+i v_k)^n\right]
+2\operatorname{Re}(p)\left[(1-w_k)^n-(1+w_k)^n\right]
}{
(1+|p|^2)\left[(1+w_k)^n+(1-w_k)^n\right]
},
\end{equation}
\begin{equation}\label{eq:bloch_update_v}
v_{k+1}=
\frac{
2\operatorname{Im}\!\left[(1+p^2)(u_k+i v_k)^n\right]
+2\operatorname{Im}(p)\left[(1+w_k)^n-(1-w_k)^n\right]
}{
(1+|p|^2)\left[(1+w_k)^n+(1-w_k)^n\right]
},
\end{equation}
\begin{equation}\label{eq:bloch_update_w}
w_{k+1}=
\frac{
(1-|p|^2)\left[(1+w_k)^n-(1-w_k)^n\right]
+4\operatorname{Re}\!\left[p(u_k+i v_k)^n\right]
}{
(1+|p|^2)\left[(1+w_k)^n+(1-w_k)^n\right]
}.
\end{equation}
For fixed $n$ and $p$, these equations define an autonomous discrete-time dynamical system on the Bloch ball. Each updated coordinate depends only on the current state, without explicit dependence on the iteration number. On the invariant unit sphere, the evolution agrees with the rational map $f_p$. Mixed-state trajectories can also change their distance from the origin, and their purity need not vary monotonically.

The maximally mixed state $\rho_{max}=I_2/2$ is fixed for every $p$: $T_0(\rho_{max})=\rho_{max}$, and unitary conjugation leaves it unchanged. Where this fixed point is attracting, its basin supplies an additional mixed-state convergence outcome. Note that being a fixed point does not imply attraction for every parameter.

The ideal reflection relations also extend to mixed inputs. The entrywise power respects both $\rho\mapsto\bar\rho$ and $\rho\mapsto X\rho X$, while $\bar U_p=U_{\bar p}$ and $XU_pX=U_{-\bar p}$. Thus these transformations conjugate the mixed-state maps at $p$, $\bar p$, and $-\bar p$, preserving purity and boundary dimension. They justify restricting the ideal mixed-state parameter scans to nonnegative real and imaginary axes.

The constant-purity surfaces considered below specify the initial states being compared; trajectories are not constrained to remain on them. We examine the convergence regions on these surfaces to determine how initial mixing affects the fractal boundary geometry.

\subsection{Spherical sections of the quasi-Julia set}\label{sec:spherical_sections}

Fractal convergence boundaries can extend into the mixed-state region of the Bloch ball. Their intersections with constant-purity surfaces allow us to examine how initial mixing affects the geometry. Following earlier work, we call quasi-Julia set the boundary between distinct pure-cycle phases (including phases of the same attracting cycle) inside the Bloch ball~\cite{robustness2023}.

We identify these regions by forward iteration of the Bloch-coordinate map. For each initial purity, we sample $10^7$ states uniformly on the corresponding sphere and record them on a $500\times500$ equal-area grid. Convergence is sought over $500$ iterations using a tolerance of $10^{-9}$, then checked over five additional cycle periods. The classification distinguishes both the attracting cycle and its phase relative to a common iteration index. Trajectories that do not satisfy the convergence test remain unresolved.
The reference pure-state cycles are independently checked against the critical orbits starting from $z=0$ and $z=\infty$.

A grid cell is marked as part of the boundary when it contains classified initial states belonging to at least two different pure-cycle outcomes. Unresolved trajectories are not assigned a separate basin label. The resulting binary grid provides a finite-resolution boundary representation, whose scaling is analyzed using the box-counting procedure introduced in Section~\ref{sec:julia_dimension}. The equal-area grid defines coordinate cells, not equal physical side lengths on the sphere. Box widths in this estimator are measured in grid cells; equality of cell areas alone does not establish an isotropic spherical covering.
This method requires distinguishable convergence outcomes, but not necessarily different attracting cycles. If only one outcome is resolved, no boundary dimension is reported.

Previous studies found that fractal boundaries survive reduced initial purity in quadratic and higher-order Hadamard protocols, with their disappearance occurring at a protocol-dependent threshold~\cite{Malachov_2019,PORTIK2022127999}. We examine this persistence across the generalized family by scanning the real and imaginary parameter axes for $n=2,3,4,$ and $5$, comparing $P=0.99$ and $P=0.98$ with the pure-state case $P=1$. Where the forward classification resolves at least two outcomes, the $P=1$ curves are the shared forward-classification reference series also used in the three-dimensional comparison below. Where no such boundary can be identified, the $P=1$ values are instead estimated by sampling the Julia set with backward iteration, as described in Section~\ref{sec:julia_dimension}. This occurs for three reasons. A single attracting fixed point can attract all initial states, leaving only one outcome. The Julia set then bounds a single basin and, at the parameters studied here, forms a dust-like set, so no $P<1$ estimate is reported although the pure-state Julia set remains well defined. Convergence can also be too slow to be resolved within $500$ iterations. Finally, the map can have no attracting cycle at all, most notably around $p=\pm i$ for even $n$.

\begin{figure}[pos=htbp]
\singlespacing
\centering
\resizebox{.85\linewidth}{!}{%
\begin{tabular}{@{}c@{\hspace{6.4mm}}c@{}}
\small Real axis, $\operatorname{Im}p=0$ & \small Imaginary axis, $\operatorname{Re}p=0$ \\[1mm]
\includegraphics{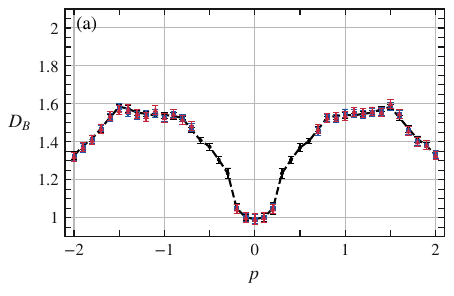} & \includegraphics{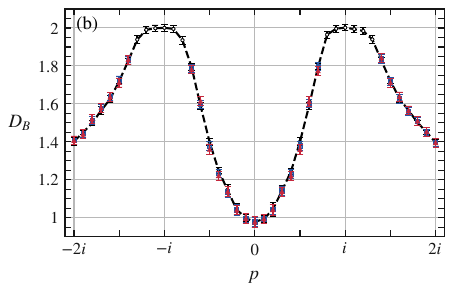} \\[1mm]
\includegraphics{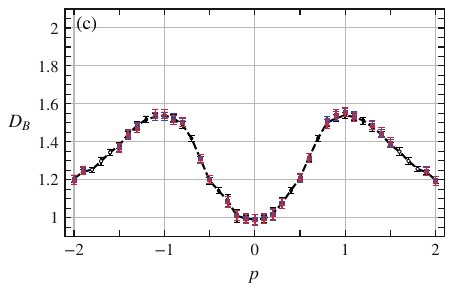} & \includegraphics{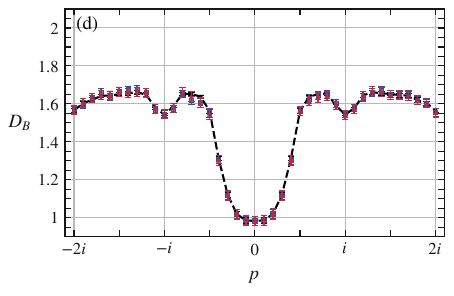} \\[1mm]
\includegraphics{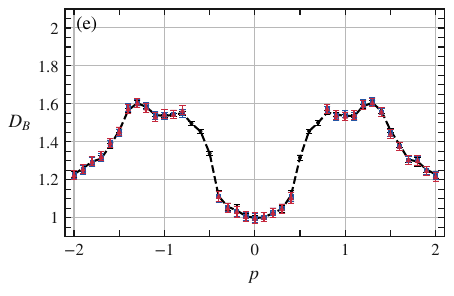} & \includegraphics{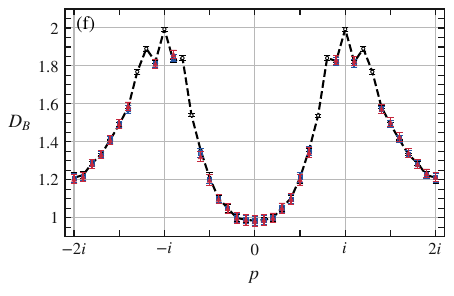} \\[1mm]
\includegraphics{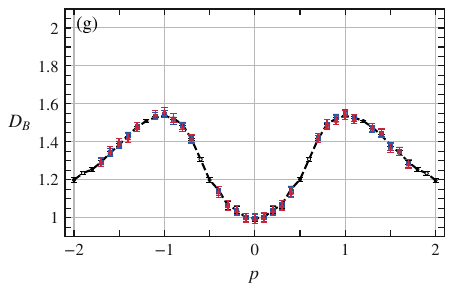} & \includegraphics{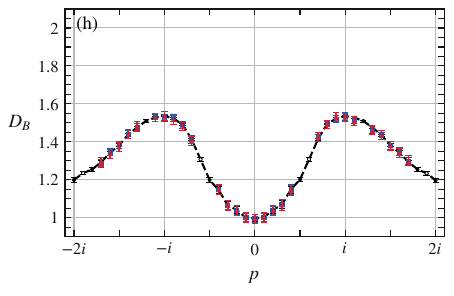}
\end{tabular}%
}
\caption{Estimated box-counting dimensions on constant-purity surfaces along the real (left column) and imaginary (right column) parameter axes. From top to bottom, the rows show orders $n=2,3,4,$ and $5$. Black circles, blue squares, and red triangles correspond to $P=1$, $0.99$, and $0.98$, respectively. Dashed black lines connect the pure-state estimates. Error bars show regression standard errors. Missing estimates indicate that the method did not resolve a boundary between distinct phase-resolved outcomes on the selected surface. Open black circles mark parameters at which no attracting cycle was found. There, the Julia set may cover the whole sphere, and the $P=1$ value obtained by backward iteration may underestimate the dimension (Section~\ref{sec:spherical_sections}). Filled black circles without $P<1$ counterparts are also obtained by backward iteration, at parameters where only one convergence outcome exists or convergence is too slow to be classified.}
\label{fig:mixed_sections_real_imag}
\end{figure}

Where boundaries are resolved, the estimated dimensions remain close across these purities and preserve the main parameter-dependent features of the pure-state profiles, as shown in Fig.~\ref{fig:mixed_sections_real_imag}. Isolated differences may reflect changes in the boundary geometry, proximity to a purity threshold, or finite-resolution effects. Missing estimates indicate that the numerical procedure did not resolve a suitable boundary; they do not establish the absence of unstable dynamics.

For parameters without attracting cycles, the box-counting estimates are limited by the finite number of sampled points and may differ from the true dimension of the Julia set. Backward iteration with randomly chosen inverse branches samples an invariant distribution whose support is the Julia set, but which is, in general, highly non-uniform. Occupancy-based box counting therefore converges to the dimension of $J(f_p)$ only in the limit of unlimited sampling, and a finite sample can underestimate it where the distribution is concentrated. This is expected to be most pronounced where the critical orbits have small Lyapunov exponents, since the sampled points then spread slowly over the set. At $p=\pm i$ for even $n$, both critical orbits land on the repelling fixed point $z=1$, so $J(f_p)=\hat{\mathbb{C}}$ and thus $D=2$, in agreement with the estimates. At the neighboring parameters, no attracting cycle of period up to $2000$ is found and the critical orbits have positive Lyapunov exponents. This suggests that the Julia set is again the whole sphere and that the estimates there underestimate its dimension. However, it cannot be excluded that very long, weakly attracting cycles exist at these parameters, whose basins are too small to be resolved numerically.

\subsection{The box-counting dimension of the quasi-Julia set}

We now extend the fractal-dimension analysis from spherical sections to the three-dimensional quasi-Julia set in the Bloch ball. The higher-dimensional calculation substantially increases the computational cost. With $r$ cells along each coordinate direction, a three-dimensional grid contains $r^3$ cells, compared with the $r^2$ scaling of the spherical sections. Memory and computation time therefore limit the accessible resolution and the precision of the dimension estimates. The box-counting procedure remains the same, but uses a three-dimensional grid to cover the sampled boundary.

To construct this cover, we sample initial states on concentric spheres, proceeding inward from the surface of the Bloch ball. Forward iteration identifies their asymptotic outcomes, distinguishing different phases of the same attracting pure-state cycle. Boundary contacts between distinct pure-state phases are detected both within each sphere and between neighboring spheres. Contacts involving only one pure-state phase and a mixed-state basin are excluded. Trajectories that cannot be classified remain unresolved. The radial increment is $\Delta r=0.01$, with $10^5$ initial states sampled on each sphere. The accumulated boundary is covered by Cartesian cubic grids with $16^3$, $32^3$, $64^3$, and $128^3$ cells. Once boundary points have been detected, inward sampling terminates after five consecutive spheres without further boundary evidence, provided that none of these spheres contains unresolved trajectories. This stopping criterion limits the computational cost, although it cannot rule out deeper boundary components. The dimension is estimated from the number of occupied boxes at different scales.

\begin{figure}[pos=htbp]
\singlespacing
\centering
\resizebox{.85\linewidth}{!}{%
\begin{tabular}{@{}c@{\hspace{6.4mm}}c@{}}
\small Real axis, $\operatorname{Im}p=0$ & \small Imaginary axis, $\operatorname{Re}p=0$ \\[1mm]
\includegraphics{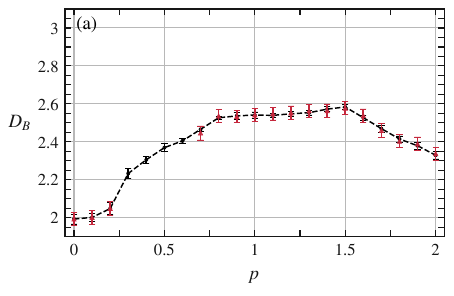} & \includegraphics{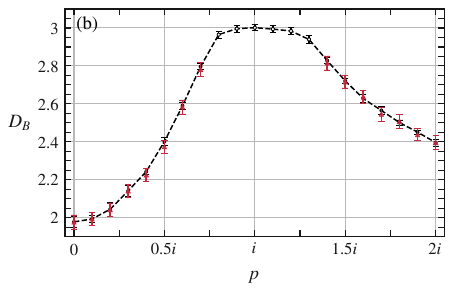} \\[1mm]
\includegraphics{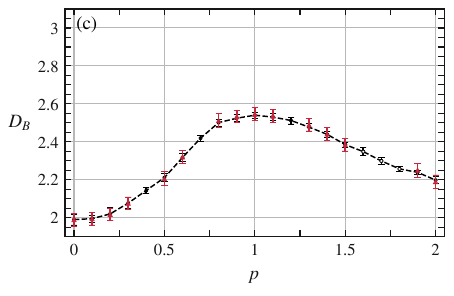} & \includegraphics{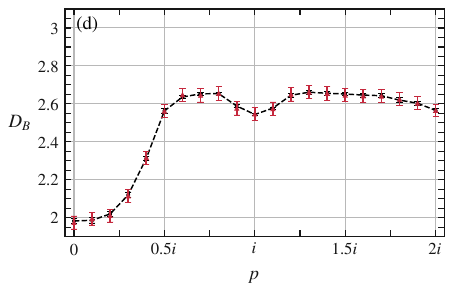} \\[1mm]
\includegraphics{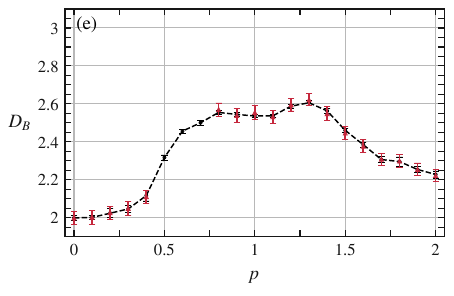} & \includegraphics{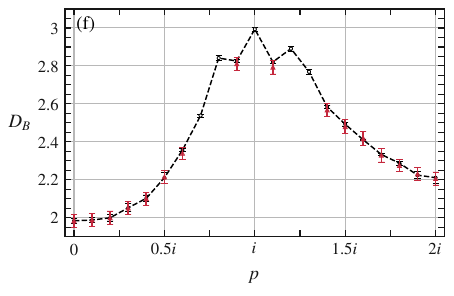} \\[1mm]
\includegraphics{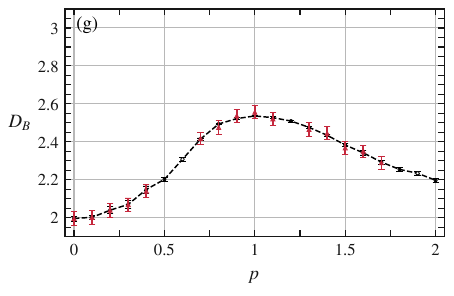} & \includegraphics{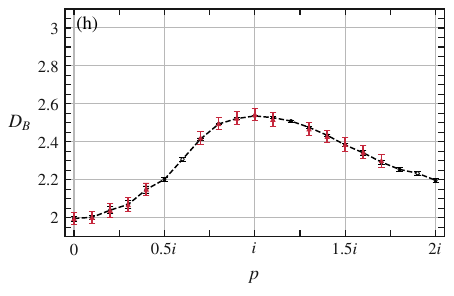}
\end{tabular}%
}
\caption{Box-counting dimension of the three-dimensional quasi-Julia set along the real (left column) and imaginary (right column) parameter axes. Rows show $n=2,3,4,$ and $5$ from top to bottom. Red triangles show $D_B^{(3\mathrm D)}$, while black circles joined by dashed lines show $1+D_B^{(2\mathrm D)}$ for the pure-state Julia set, $P=1$. Error bars show regression standard errors. Missing red triangles indicate that no three-dimensional boundary between distinct pure-state phases was resolved numerically.}
\label{fig:3D_real_imag}
\end{figure}

The reflection symmetries of the dynamics allow us to restrict the parameter scans to the nonnegative real and imaginary axes. Figure~\ref{fig:3D_real_imag} shows the resulting box-counting dimensions for protocol orders $n=2,3,4,$ and $5$. Most curves follow the overall behavior found for the pure-state Julia sets: relatively low dimensions at small $|p|$, a pronounced increase toward intermediate parameter values, and a subsequent decrease toward larger $|p|$. Broad maxima or plateaus occur around $|p|\approx1$, although their positions and widths depend on the order and the direction of the parameter scan. Additional peaks and depressions are also present, so the profiles do not all follow a single monotonic rise and decay. These similarities show that the main parameter-dependent trends persist in the three-dimensional mixed-state setting.

To compare the dimensions directly, the figure also includes the corresponding pure-state Julia-set dimensions shifted upward by one. Over much of the parameter range where both estimates are available, the curves approximately coincide, giving
\begin{equation}\label{eq:dimension_shift}
\Delta D_B\equiv D_B^{(3\mathrm D)}-D_B^{(2\mathrm D)}\approx1.
\end{equation}
Here, $D_B^{(3\mathrm D)}$ refers to the three-dimensional quasi-Julia boundary, while $D_B^{(2\mathrm D)}$ refers specifically to the pure-state Julia set at $P=1$. The comparison does not imply the same difference for every constant-purity section.

The observed difference admits a geometric interpretation through the usual codimension-one slicing picture: intersecting a fractal in three-dimensional space with a two-dimensional surface can reduce its dimension by one~\cite{Falconer2003}. However, the pure-state sphere is a distinguished boundary of the Bloch ball rather than a generic section. The approximate relation above is therefore a numerical observation for the studied protocols, not a general identity for box-counting dimensions.

\subsection{Purity-driven phase transition}

Reducing the initial purity can lead to an abrupt loss of fractal structure in the boundaries between convergence regions. Previous studies of quadratic and higher-order Hadamard protocols found that the estimated boundary dimension remains approximately constant above a critical purity $P_{\mathrm c}$, then falls to approximately one where a smooth boundary survives below the transition~\cite{Malachov_2019,PORTIK2022127999}. For the Hadamard family, the critical purity increases with the protocol order and approaches one at large $n$, indicating a decreasing tolerance of the fractal structure to preparation noise~\cite{PORTIK2022127999}. For the earlier ideal Hadamard protocols, numerical thresholds agree with the purity of a mixed repelling fixed point on the common boundary of the phase-resolved convergence regions~\cite{Malachov_2019,PORTIK2022127999}. We do not establish this correspondence for every parameter in the generalized family.

The three-dimensional quasi-Julia structure provides a geometric interpretation of this transition. The boundaries observed on constant-purity spheres are sections of the structure inside the Bloch ball. As the initial purity decreases, the sampling sphere moves inward and probes progressively more mixed states. Fractal sections can persist while this sphere intersects the quasi-Julia structure, accounting for the survival of complex boundaries over a finite purity range. The nearly unchanged sectional dimensions reported above are a numerical property of the investigated cases; they do not follow merely from the dimension of the sampling surface. Planar sections can also reveal this geometry, although sections selected by a symmetry may display different features.

The critical purity marks the lower extent in purity of the fractal structure. Below this value, the constant-purity sphere no longer intersects that structure. Earlier Hadamard calculations found that a smooth convergence boundary can remain, with sectional dimension near one~\cite{Malachov_2019,PORTIK2022127999}. Such a boundary may separate a pure phase from a mixed attractor. The pure-phase classifier used here excludes these contacts and therefore need not detect that residual boundary. If fewer than two pure phases are resolved, no quasi-Julia boundary dimension is reported. The abrupt change in the measured dimension can thus be understood as the loss of the fractal intersection, without requiring every remaining boundary to disappear. Finite sampling limits how precisely the threshold can be located and whether small boundary components can be resolved.

Above the threshold, preparation noise therefore need not remove the sensitivity associated with fractal convergence boundaries. Nearby initial states lying on opposite sides of such a boundary can still approach different asymptotic outcomes, including different phases of the same attracting cycle. Moderate mixing preserves this source of sensitivity wherever the fractal boundary remains accessible, while the purity-driven transition limits the range of initial states for which it is observed.

\section{The impact of gate noise}
\label{sec:gate_noise}

Preparation noise enters through the initial state, whereas gate errors modify the transformation repeated during the protocol. We now examine how stochastic noise and fixed coherent errors affect the attractors, convergence regions, and fractal boundaries. Earlier work considered coherent miscalibration of the Hadamard gate~\cite{robustness2023}. The competition between repeated purification and decoherence has also been studied through fractal boundaries in mixed-state space~\cite{viennot2020competition}. Here, we study effective errors in the pre-measurement evolution of the generalized protocol, keeping the measurement, postselection, and final gate $U_p$ ideal.

\subsection{Description of gate errors}

The input to each round consists of $n$ uncorrelated qubits in the same state, $\Sigma_k=\rho_k^{\otimes n}$. A faulty unitary operation can be represented by its ideal channel followed by an error channel. A fixed unitary error describes a coherent miscalibration, while a general quantum channel can describe stochastic processes such as random Pauli errors, depolarization, or damping~\cite{Wallman2015}. Coherent and stochastic descriptions are not generally interchangeable, as illustrated by comparisons of their effects in quantum error correction~\cite{042338,aa9a06}. A channel representing averaged stochastic noise need not vary between iterations. Imperfect measurement and postselection would instead require specifying the corresponding accepted-outcome operation, which can change both the accepted state and its acceptance probability.

Rather than assigning a separate error parameter to every gate, we introduce an effective $n$-qubit channel $\mathcal C_k$ acting after the ideal core evolution and before postselection,
\begin{equation}\label{eq:E_effective}
\mathcal E_k=\mathcal C_k\circ\mathcal E_0.
\end{equation}
This placement defines the models considered below without asserting that the physical errors occur at a single location in the circuit. The unnormalized accepted output, its success probability, and the conditional update are
\begin{align}\label{eq:effective_conditional_map}
N_{\mathcal C,k}(\rho_k)
&=\operatorname{Tr}_{2,\ldots,n}
\!\left[\Pi_0\mathcal E_k(\Sigma_k)\Pi_0\right],\nonumber\\
D_{\mathcal C,k}(\rho_k)
&=\operatorname{Tr}N_{\mathcal C,k}(\rho_k),\nonumber\\
F_{\mathcal C,k}(\rho_k)
&=U_p\frac{N_{\mathcal C,k}(\rho_k)}{D_{\mathcal C,k}(\rho_k)}U_p^\dagger.
\end{align}
The joint-register channel is linear, but restricting its input to $\rho_k^{\otimes n}$ and normalizing the accepted output gives a nonlinear single-qubit map. A fixed channel defines an autonomous map; an iteration-dependent channel gives a sequence of maps.

To separate coherent and stochastic contributions in a representative weak-error model, we write
\begin{equation}\label{eq:E_composed}
\mathcal E_{\varepsilon,\eta,k}
=\Lambda_{\eta,k}\circ\mathcal U_\varepsilon\circ\mathcal E_0,
\qquad
\mathcal U_\varepsilon(X)=V_\varepsilon X V_\varepsilon^\dagger,
\qquad V_\varepsilon=e^{-i\varepsilon H_{\mathrm{err}}},
\end{equation}
where $H_{\mathrm{err}}$ is a fixed Hermitian generator and $\Lambda_{\eta,k}$ is a completely positive trace-preserving channel. The parameters $\varepsilon$ and $\eta$ characterize the coherent and stochastic error strengths. This factorization specifies an effective model rather than a unique decomposition of an arbitrary error channel.

The channel order generally matters because $\Lambda_{\eta,k}$ and $\mathcal U_\varepsilon$ need not commute. If both approach the identity smoothly, reversing their order changes terms starting at order $\varepsilon\eta$, leaving the first-order expansion unchanged. The global depolarizing channel considered below commutes exactly with every unitary channel, so its order relative to the coherent error is immaterial even beyond first order. These statements concern the two error channels, whose placement relative to postselection remains fixed.

Expanding near $\varepsilon=\eta=0$ gives
\begin{align}\label{eq:channel_expansion}
\mathcal U_\varepsilon(X)
&=X+\varepsilon\Delta\mathcal U(X)+\mathcal O(\varepsilon^2),\nonumber\\
\Lambda_{\eta,k}(X)
&=X+\eta\Delta\Lambda_k(X)+\mathcal O(\eta^2),
\end{align}
where $\Delta\mathcal U(X)=-i[H_{\mathrm{err}},X]$. Thus,
\begin{equation}\label{eq:E_expand}
\mathcal E_{\varepsilon,\eta,k}
=\mathcal E_0+\varepsilon\Delta\mathcal E_{\mathrm{sys}}
+\eta\Delta\mathcal E_{\mathrm{stoch},k}
+\mathcal O(\varepsilon^2,\eta^2,\varepsilon\eta),
\end{equation}
with $\Delta\mathcal E_{\mathrm{sys}}=\Delta\mathcal U\circ\mathcal E_0$ and $\Delta\mathcal E_{\mathrm{stoch},k}=\Delta\Lambda_k\circ\mathcal E_0$. For either contribution, we define
\begin{equation}\label{eq:accepted_corrections}
\Delta N_x(\rho_k)
=\operatorname{Tr}_{2,\ldots,n}
\!\left[\Pi_0\Delta\mathcal E_x(\Sigma_k)\Pi_0\right],
\qquad
\Delta D_x(\rho_k)=\operatorname{Tr}\Delta N_x(\rho_k),
\end{equation}
where $x\in\{\mathrm{sys},\mathrm{stoch}\}$ and the index $k$ is understood for an iteration-dependent stochastic channel. The accepted output and success probability inherit these expansions,
\begin{align}\label{eq:ND_expansion}
N_{\varepsilon,\eta,k}
&=N_0+\varepsilon\Delta N_{\mathrm{sys}}
+\eta\Delta N_{\mathrm{stoch},k}
+\mathcal O(\varepsilon^2,\eta^2,\varepsilon\eta),\nonumber\\
D_{\varepsilon,\eta,k}
&=D_0+\varepsilon\Delta D_{\mathrm{sys}}
+\eta\Delta D_{\mathrm{stoch},k}
+\mathcal O(\varepsilon^2,\eta^2,\varepsilon\eta).
\end{align}
Expanding their ratio yields
\begin{equation}\label{eq:F_first_order}
F_{\varepsilon,\eta,k}(\rho_k)
=F_0(\rho_k)+\varepsilon B_{\mathrm{sys}}(\rho_k)
+\eta B_{\mathrm{stoch},k}(\rho_k)
+\mathcal O(\varepsilon^2,\eta^2,\varepsilon\eta),
\end{equation}
where
\begin{equation}\label{eq:B_general}
B_x(\rho_k)
=U_p\frac{\Delta N_x(\rho_k)-T_0(\rho_k)\Delta D_x(\rho_k)}
{D_0(\rho_k)}U_p^\dagger.
\end{equation}
The correction depends on the state even for a fixed error channel, with additional dependence introduced by the postselection normalization. The expansion requires the change in the success probability to be small relative to $D_0$. Hermiticity and normalization of the exact conditional output imply
\begin{equation}\label{eq:B_traceless_hermitian}
B_x(\rho)^\dagger=B_x(\rho),
\qquad \operatorname{Tr}B_x(\rho)=0.
\end{equation}
These conditions do not guarantee positivity of a truncated approximation for arbitrary error strengths.

A fixed coherent error changes the same nonlinear map at every iteration and can shift attractors, alter their stability, and deform convergence regions. An averaged stochastic channel also defines a deterministic update when its parameters are fixed, generally introducing mixing and a nonzero first-order correction. Fluctuations of individual noise histories require a separate stochastic description. In particular, stochastic noise does not by itself imply a zero-mean perturbation, diffusion, or a steady-state noise floor.

Postselection also changes the relative weights of noise histories. If a realization $r$ occurs with probability $q_r$, succeeds with probability $D_r$, and yields the normalized accepted state $\sigma_r$, the accepted ensemble is described by
\begin{equation}\label{eq:postselection_noise_weights}
\sigma_{\mathrm{accepted}}
=\frac{\sum_r q_rD_r\sigma_r}{\sum_r q_rD_r}.
\end{equation}
It is generally different from $\sum_rq_r\sigma_r$. Applying the averaged channel before selecting and normalizing the accepted output incorporates this reweighting. The calculations below use this density-matrix evolution, without resolving individual noise histories, and assume independent inputs in each round.

\subsection{Depolarizing noise}

We first consider an $n$-qubit depolarizing channel acting once per iteration,
\begin{equation}\label{eq:global_depolarization}
\Lambda_\eta(\Sigma)=(1-\eta)\Sigma+\eta\frac{I_{2^n}}{2^n},
\qquad 0\leq\eta\leq1,
\end{equation}
for a normalized joint state $\Sigma$. Here, $\eta$ is a global replacement probability per round, rather than an error probability assigned separately to each CNOT. Independent gate-local depolarizing channels would generally produce a different conditional map.

With $\mathcal E_\eta=\Lambda_\eta\circ\mathcal E_0$, linearity gives
\begin{align}\label{eq:depol_N}
N_\eta(\rho)
&=(1-\eta)N_0(\rho)
+\eta\operatorname{Tr}_{2,\ldots,n}
\!\left[\Pi_0\frac{I_{2^n}}{2^n}\Pi_0\right]\nonumber\\
&=(1-\eta)N_0(\rho)+\eta\frac{I_2}{2^n}.
\end{align}
The maximally mixed contribution is restricted to the selected ancilla state, giving the success probability
\begin{equation}\label{eq:depol_D}
D_\eta(\rho)=(1-\eta)D_0(\rho)+\eta2^{1-n}.
\end{equation}
Using $N_0=D_0T_0$, the normalized accepted state can be written exactly as
\begin{equation}\label{eq:Tq_exact_mixture}
T_\eta(\rho)
=\bigl(1-\omega_\eta(\rho)\bigr)T_0(\rho)
+\omega_\eta(\rho)\frac{I_2}{2},
\qquad \text{with} \qquad
\omega_\eta(\rho)
=\frac{\eta2^{1-n}}{(1-\eta)D_0(\rho)+\eta2^{1-n}}.
\end{equation}
Including the final unitary gives
\begin{equation}\label{eq:F_depol_exact}
F_\eta(\rho)
=\bigl(1-\omega_\eta(\rho)\bigr)F_0(\rho)
+\omega_\eta(\rho)\frac{I_2}{2}.
\end{equation}
Postselection therefore converts a state-independent global replacement probability into a state-dependent mixing weight. For nonzero $\eta$, the output is drawn toward the Bloch-ball interior relative to the ideal output. This does not imply that purity decreases monotonically along a trajectory, since the ideal nonlinear step can itself purify a mixed input. Unlike preparation noise, this mixing acts in every round.

\subsection{Coherent gate errors}

We model a fixed coherent error in the core evolution by a residual interaction on the qubit pairs connected by the CNOT network. The effective unitary acts after the ideal core and before postselection,
\begin{equation}\label{eq:coherent_error_channel}
\mathcal E_\varepsilon=\mathcal U_\varepsilon\circ\mathcal E_0,
\qquad V_\varepsilon=e^{-i\varepsilon H}.
\end{equation}
This describes a repeatable interaction during one nonlinear round, without specifying a microscopic error model for each individual CNOT.

For the analytical example, we choose
\begin{equation}\label{eq:ZZ_generator}
H_{ZZ}=\sum_{j=2}^n Z_1Z_j.
\end{equation}
This generator is diagonal in the computational basis, so $V_\varepsilon$ commutes with $\Pi_0$. Consequently,
\begin{align}\label{eq:Neps_commuted}
N_\varepsilon(\rho_k)
&=\operatorname{Tr}_{2,\ldots,n}
\!\left[\Pi_0 V_\varepsilon\mathcal E_0(\Sigma_k)V_\varepsilon^\dagger\Pi_0\right]\nonumber\\
&=\operatorname{Tr}_{2,\ldots,n}
\!\left[V_\varepsilon\Pi_0\mathcal E_0(\Sigma_k)\Pi_0V_\varepsilon^\dagger\right].
\end{align}
On the selected ancilla subspace, every target has $Z$ eigenvalue $+1$. Hence,
\begin{equation}\label{eq:ueps}
\Pi_0V_\varepsilon\Pi_0
=u_\varepsilon\otimes|0\dots0\rangle\langle0\dots0|,
\qquad \text{where} \qquad
u_\varepsilon=e^{-i\varepsilon(n-1)Z}.
\end{equation}
The accepted output and its trace are therefore
\begin{equation}\label{eq:ZZ_accepted_output}
N_\varepsilon(\rho)=u_\varepsilon N_0(\rho)u_\varepsilon^\dagger,
\qquad D_\varepsilon(\rho)=D_0(\rho).
\end{equation}
The success probability is unchanged for the same input state, although noisy and ideal trajectories can visit different states and thus have different success probabilities in later rounds. The exact conditional map is
\begin{equation}\label{eq:Feps_exact}
F_\varepsilon(\rho)
=U_pu_\varepsilon T_0(\rho)u_\varepsilon^\dagger U_p^\dagger.
\end{equation}
The coherent correction rotates the ideal intermediate output without changing its purity. Pure inputs remain pure, but the complete nonlinear step need not conserve the purity of mixed inputs. The Bloch rotation angle is $2\varepsilon(n-1)$, so holding $\varepsilon$ fixed increases the effective rotation with the protocol order.

For pure states, Eq.~\eqref{eq:Feps_exact} gives
\begin{equation}\label{eq:ZZ_pure_map}
f_{p,\varepsilon}^{ZZ}(z)=
\frac{e^{2i(n-1)\varepsilon}z^n-\bar p}{1+p e^{2i(n-1)\varepsilon}z^n}.
\end{equation}
Writing $R_\varepsilon(z)=e^{-2i\varepsilon}z$, direct substitution yields
\begin{equation}\label{eq:ZZ_conjugacy}
R_\varepsilon^{-1}\circ f_{p,\varepsilon}^{ZZ}\circ R_\varepsilon
=f_{p e^{-2i\varepsilon}},
\qquad
D_B^{ZZ}(p,\varepsilon)=D_B^{\mathrm{ideal}}(p e^{-2i\varepsilon}).
\end{equation}
Thus $ZZ$ errors rotate the ideal parameter rather than defining a separate family of Julia-set dimensions. This also holds for mixed inputs. If $\mathcal C_\theta(\rho)=e^{i\theta Z}\rho e^{-i\theta Z}$, entrywise powering gives $T_0\circ\mathcal C_\theta=\mathcal C_{n\theta}\circ T_0$, and
\begin{equation}\label{eq:ZZ_mixed_conjugacy}
\mathcal C_\varepsilon^{-1}\circ F_{\varepsilon,p}^{ZZ}\circ\mathcal C_\varepsilon
=F_{0,p e^{-2i\varepsilon}}.
\end{equation}
Here, $F_{\varepsilon,p}^{ZZ}$ denotes the map of Eq.~\eqref{eq:Feps_exact} and $F_{0,p}$ the ideal map of Eq.~\eqref{eq:F0_def}, both at parameter $p$. The coordinate rotation preserves each purity sphere and its boundary dimension. Although the intermediate output rotation grows with $n$, the conjugate parameter rotation is $2\varepsilon$, independent of order. For noisy parameter reflections, $\rho\mapsto X\rho X$ maps $(p,\varepsilon)$ to $(-\bar p,-\varepsilon)$; it does not establish a reflection symmetry at fixed nonzero strength. The noisy scans therefore include both signs of the parameter axes.

For comparison, we also consider
\begin{equation}\label{eq:XX_YY_generators}
H_{XX}=\sum_{j=2}^n X_1X_j,
\qquad
H_{YY}=\sum_{j=2}^n Y_1Y_j,
\end{equation}
using the same unitary parametrization. These generators do not generally commute with the postselection projector, and the corresponding accepted maps are evaluated numerically. The unchanged-success-probability result for $ZZ$ cannot be extended to them. Pure inputs nevertheless yield pure accepted outputs whenever acceptance has nonzero probability, because the joint evolution is unitary and all measured targets are projected onto a specified pure state.

\subsection{Comparison of the noisy dynamics}

\begin{figure}[pos=H]
\singlespacing
\centering
\begin{tabular}{@{}c@{}}
\includegraphics{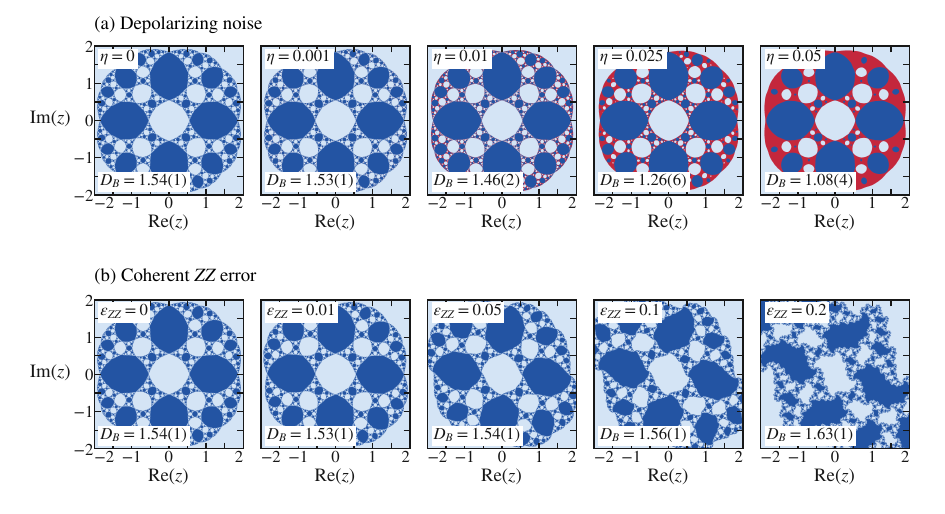}
\end{tabular}
\caption{Convergence regions for pure initial states in the quadratic $p=1$ protocol. From left to right, row (a) shows depolarization probabilities $\eta=0$, $0.001$, $0.01$, $0.025$, and $0.05$, while row (b) shows coherent $ZZ$ strengths $\varepsilon_{ZZ}=0$, $0.01$, $0.05$, $0.1$, and $0.2$. Light and dark blue distinguish the two phases of convergence to the attracting cycle, while red denotes convergence to the maximally mixed state. The error strength is given in the upper-left corner of each panel, and the estimated box-counting dimension is given in the lower-left corner. Regression standard errors are rounded upward to $0.01$ and given in parentheses in units of the last displayed decimal place.}
\label{fig:basins_depol_n2_p1}
\label{fig:basins_ZZ_n2_p1}
\label{fig:basins_noise_comparison_n2_p1}
\end{figure}

We compare the error models through their convergence regions and the estimated box-counting dimensions of the boundaries. For the convergence-region plots, we use forward iteration to classify convergence to the attractors of the noisy map and apply the boundary-counting procedure described for preparation noise. Different phases of an attracting cycle remain separate outcomes, and trajectories that fail the convergence test remain unresolved. In these basin plots, the initial states are pure; under depolarization, subsequent states can lie inside the Bloch ball. Figure~\ref{fig:basins_noise_comparison_n2_p1} illustrates the quadratic $p=1$ protocol.

\begin{figure}[pos=htbp]
\singlespacing
\centering
\resizebox{.85\linewidth}{!}{%
\begin{tabular}{@{}c@{\hspace{6.4mm}}c@{}}
\small Real axis, $\operatorname{Im}p=0$ & \small Imaginary axis, $\operatorname{Re}p=0$ \\[1mm]
\includegraphics{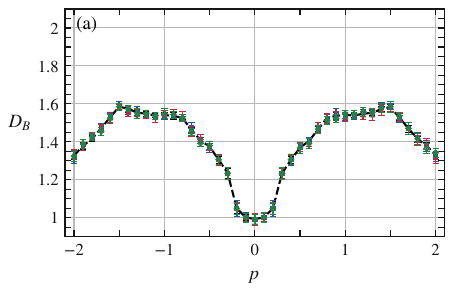} & \includegraphics{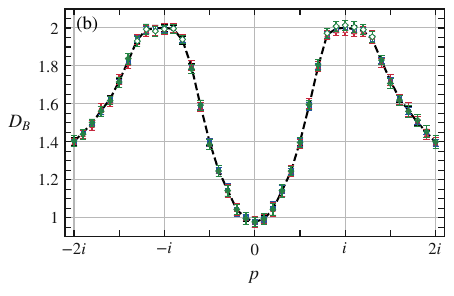} \\[1mm]
\includegraphics{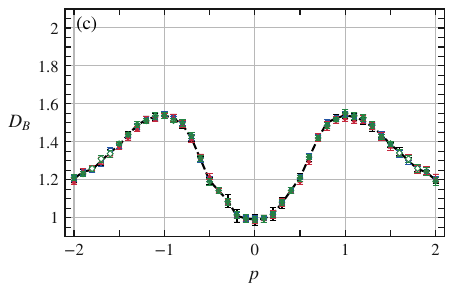} & \includegraphics{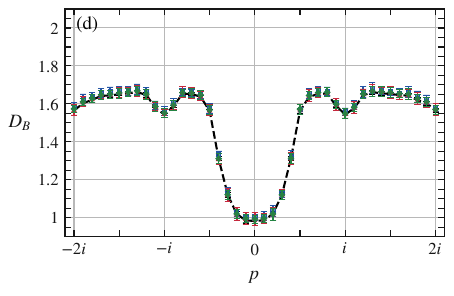} \\[1mm]
\includegraphics{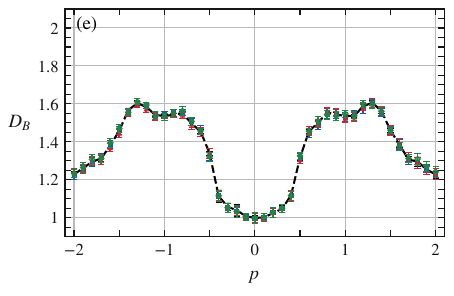} & \includegraphics{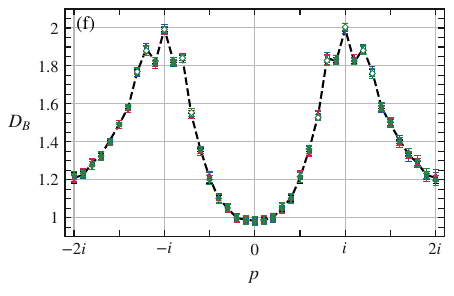} \\[1mm]
\includegraphics{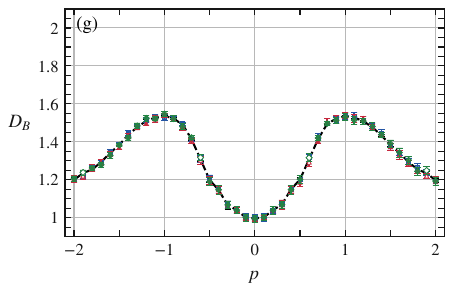} & \includegraphics{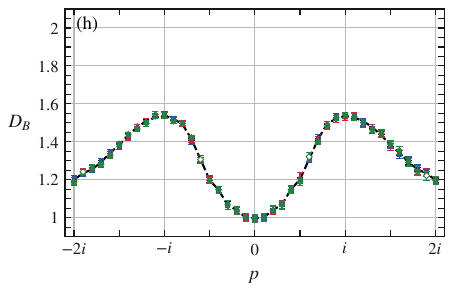}
\end{tabular}%
}
\caption{Estimated box-counting dimensions along the real (left column) and imaginary (right column) parameter axes under weak coherent gate errors. From top to bottom, the rows show orders $n=2,3,4,$ and $5$. Black circles show the ideal protocol, blue squares the $ZZ$ model, red triangles the $XX$ model, and green diamonds the $YY$ model, with $\varepsilon=0.01$ for each coherent error model. Dashed black lines connect the ideal estimates. Error bars show regression standard errors. Missing estimates indicate that no boundary suitable for fitting was resolved. Open symbols mark parameters at which the ideal map has no attracting cycle, where the estimates may underestimate the dimension (Section~\ref{sec:spherical_sections}).}
\label{fig:box_counting_systematic_real_imag}
\end{figure}

\begin{figure}[pos=htbp]
\singlespacing
\centering
\resizebox{.85\linewidth}{!}{%
\begin{tabular}{@{}c@{\hspace{6.4mm}}c@{}}
\small Real axis, $\operatorname{Im}p=0$ & \small Imaginary axis, $\operatorname{Re}p=0$ \\[1mm]
\includegraphics{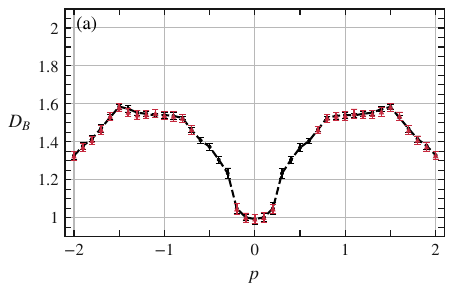} & \includegraphics{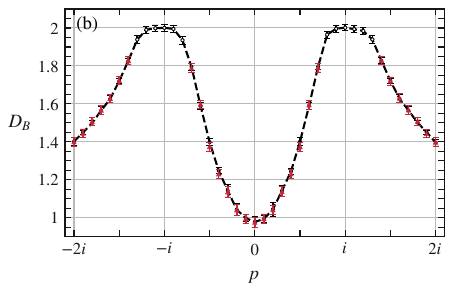} \\[1mm]
\includegraphics{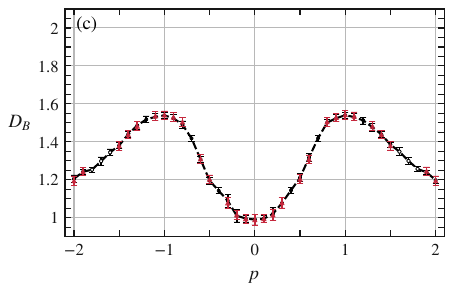} & \includegraphics{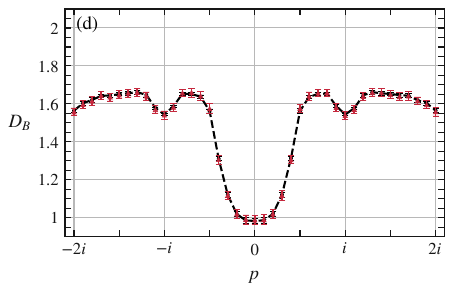} \\[1mm]
\includegraphics{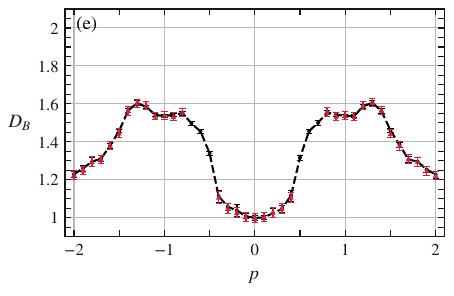} & \includegraphics{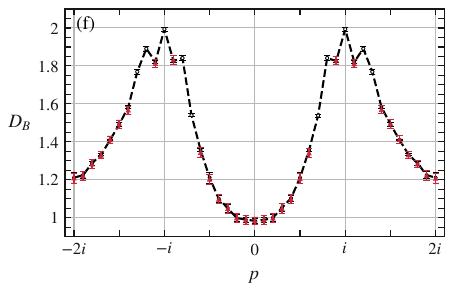} \\[1mm]
\includegraphics{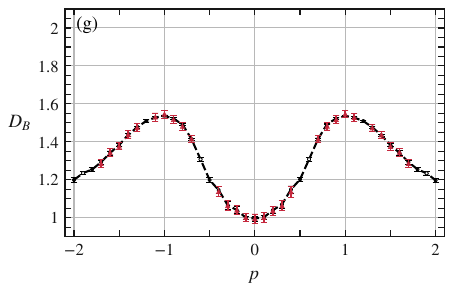} & \includegraphics{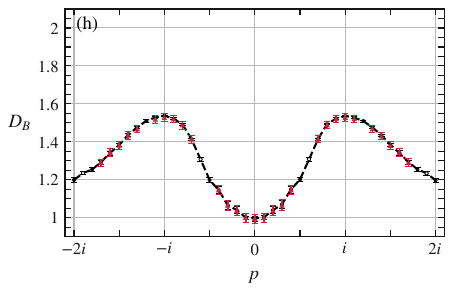}
\end{tabular}%
}
\caption{Estimated box-counting dimensions along the real (left column) and imaginary (right column) parameter axes under weak depolarizing noise. From top to bottom, the rows show orders $n=2,3,4,$ and $5$. Black circles show the ideal protocol, and red triangles show the depolarizing model at $\eta=10^{-3}$. Dashed black lines connect the ideal estimates. Error bars show regression standard errors. Missing estimates indicate parameters for which the convergence-based method did not resolve a boundary suitable for dimension estimation.}
\label{fig:box_counting_depol_real_imag}
\end{figure}

These basin calculations use a $1000\times1000$ stereographic grid over $\operatorname{Re}z,\operatorname{Im}z\in[-2,2]$ and $100$ forward iterations. Convergence is classified using a componentwise Bloch-coordinate tolerance of $10^{-3}$. For these plots, the dimension fit uses box widths of $2$, $4$, $8$, $16$, and $32$ grid pixels, and the reported uncertainties are the regression standard errors. Depolarization shifts the attracting cycle into the Bloch-ball interior. The earlier quasi-Julia definition based on pure attracting outcomes therefore does not apply directly, although convergence boundaries can remain fractal. At weak depolarization, their estimated dimensions remain close to the ideal value within numerical accuracy. The larger depolarizing noise probabilities displayed in the upper row produce visible basin deformations and lower dimension estimates.

The coherent $ZZ$ error instead changes the positions of the pure attracting states without introducing mixing. Small strengths preserve the qualitative convergence structure and leave the estimated boundary dimension close to its ideal value. The larger strengths displayed in the lower row produce stronger deformations and higher dimension estimates. These observations concern the sampled strengths and do not establish a universal monotonic dependence on coherent or depolarizing error strength.

To examine the persistence of the parameter-dependent trends, we also scan the real and imaginary axes of $p$ for orders $n=2,3,4,$ and $5$. Figure~\ref{fig:box_counting_systematic_real_imag} compares the ideal results with the $ZZ$, $XX$, and $YY$ models at $\varepsilon=0.01$. Figure~\ref{fig:box_counting_depol_real_imag} shows the corresponding comparison for depolarization at $\eta=10^{-3}$. Coherent errors map pure inputs to pure outputs, so for the $ZZ$, $XX$, and $YY$ models the pure-state dynamics is still described by a rational map of degree at most $n$. We therefore sample the corresponding Julia sets by backward iteration in the same way as for the ideal map, with the inverse branches derived for each error model. Depolarization, in contrast, moves the states into the interior of the Bloch ball. Its dimensions are estimated from forward classification only, and parameters without a resolved boundary are left without an estimate. These are comparisons at fixed effective model parameters, rather than matched physical gate infidelities across orders or error types.

Where both ideal and noisy boundaries are resolved, the weak coherent errors preserve the main parameter-dependent profiles. Differences in the estimated dimensions are generally comparable to or smaller than the regression uncertainties. For $ZZ$, Eq.~\eqref{eq:ZZ_conjugacy} identifies the comparison as a displacement through the ideal parameter landscape; the small displacement need not imply exact equality of dimensions. The $XX$ and $YY$ results are separate numerical comparisons, for which no corresponding conjugacy is asserted. The detailed convergence regions can still shift or deform, so this agreement concerns the measured scaling of the boundary rather than an unchanged geometry. Similar agreement is found for the weak depolarizing channel. These comparisons do not establish exact invariance of the fractal dimension or exclude changes below the accessible resolution.

\section{Implementation and diagnostic applications}
\label{sec:implementation}

The preceding results describe the dynamics through convergence regions and their fractal boundaries. Reconstructing these structures experimentally would require sampling many initial states and performing enough successful iterations to distinguish their outcomes. Postselection makes this demanding. Related nonlinear transformations have been proposed for cavity quantum electrodynamics~\cite{torres2017measurement} and demonstrated in photonic experiments~\cite{zhu2019experimental,qu2021observation}. A more accessible experiment for the present protocol follows selected reference states whose ideal evolution is known, using their deviations to test the implementation.

Related postselected protocols have already been investigated for quantum-device testing and benchmarking~\cite{2210.09674,Ortega2024Benchmark}. Here, the fixed points of the generalized map provide reference states with adjustable local sensitivity. Such a test probes the combined implementation of preparation, entangling gates, measurement, and the final single-qubit transformation. Its outcome characterizes this protocol family rather than the computational performance of a device in general, consistent with the distinction between the capabilities measured by different benchmarks~\cite{Proctor2025BenchmarkingQCs}.

\subsection{Postselection and resource requirements}

For an ideal round acting on qubits in state $\rho$, the acceptance probability is
\begin{equation}\label{eq:implementation_success}
q(\rho)=\operatorname{Tr}\!\left[
\Pi_0\mathcal E_0(\rho^{\otimes n})\Pi_0
\right]
=\left(\frac{1+w}{2}\right)^n
+\left(\frac{1-w}{2}\right)^n.
\end{equation}
Here $q(\rho)=D_0(\rho)$, relating the resource notation to the single-round acceptance defined above. The final gate $U_p$ does not affect acceptance in that round. However, it changes the state supplied to subsequent rounds and therefore influences their probabilities. For a pure input,
\begin{equation}\label{eq:pure_input_success}
q(z)=\frac{1+|z|^{2n}}{(1+|z|^2)^n}.
\end{equation}
This probability is independent of the phase of $z$, reaches one at the computational basis states, and has minimum $2^{1-n}$ for equal populations. Restricting the input to real amplitudes simplifies the analysis but does not increase acceptance relative to a complex state with the same $|z|$. Moreover, basis inputs are not generally fixed states: the final unitary can transform them into superpositions.

\begin{figure}[pos=htbp]
\singlespacing
\centering
\begin{tabular}{@{}c@{\hspace{6.4mm}}c@{}}
\includegraphics{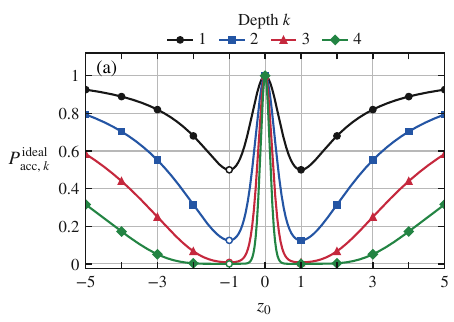} & \includegraphics{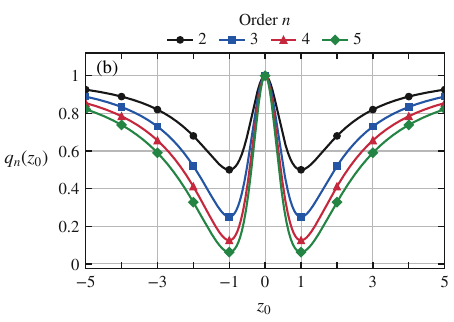}
\end{tabular}
\caption{Ideal postselection acceptance for pure inputs with real parameter $z_0$. (a) Complete-tree acceptance $P_{\mathrm{acc},k}^{\mathrm{ideal}}=q_2(z_0)^{2^k-1}$ for quadratic protocols at depths $k=1,2,3,4$, with each protocol chosen to fix $z_0$. (b) Single-round acceptance $q_n(z_0)=\frac{1+z_0^{2n}}{(1+z_0^2)^n}$ for orders $n=2,3,4,5$. Filled markers show calculated probabilities. Open circles in (a) show the probabilities approached as $z_0$ gets closer to $-1$. At $z_0=-1$, no finite protocol parameter $p$ exists. Single-round acceptance in (b) remains defined there.}
\label{fig:implementation_success}
\end{figure}

A direct implementation of $k$ iterations has an $n$-ary tree structure. It starts with $n^k$ independently prepared qubits and contains $(n^k-1)/(n-1)$ elementary nonlinear steps. With the CNOT-star implementation, the complete tree contains $n^k-1$ CNOT gates. If every postselection must succeed, its ideal acceptance probability is
\begin{equation}\label{eq:tree_acceptance_general}
P_{\mathrm{acc},k}^{\mathrm{ideal}}
=\prod_{j=0}^{k-1}
q(\rho_j)^{\,n^{k-1-j}},
\qquad \rho_j=F_0^{\,j}(\rho_0).
\end{equation}
At a fixed point, $q(\rho_j)=q_0$, giving
\begin{equation}\label{eq:tree_acceptance_fixed}
P_{\mathrm{acc},k}^{\mathrm{ideal}}
=q_0^{(n^k-1)/(n-1)}.
\end{equation}
Figure~\ref{fig:implementation_success} shows the single-round and complete-tree acceptance probabilities. Even moderately high single-step acceptance can therefore become restrictive as the tree grows. For independent complete-tree trials, obtaining a prescribed number $N_{\mathrm{acc}}$ of accepted outputs requires an expected $N_{\mathrm{acc}}/P_{\mathrm{acc},k}$ raw trials. The ideal probability supplies a resource estimate, while an experiment must use its actual acceptance rate to determine the sampling budget.

Mid-circuit measurement and reset allow measured qubits to be reused. In an ideal depth-first schedule, the required working register can be reduced to $1+(n-1)k$ qubits by storing accepted outputs while preparing the remaining inputs. This reduces simultaneous qubit use, but leaves the preparation and gate workload large and introduces waiting times for stored states. Connectivity, reset fidelity, and storage noise then become part of the implementation cost. When measured targets are not reused, their measurements may instead be deferred until the end of a complete-tree trial. Failed postselection generally changes the surviving qubit, so resetting the measured targets does not restore the input. A new attempt therefore requires freshly prepared inputs.

\subsection{Reference states with adjustable sensitivity}

To obtain a simple ideal reference, we choose a pure state and construct a protocol that leaves it fixed. For real $z_0$, the condition $f_p(z_0)=z_0$ gives
\begin{equation}\label{eq:reference_parameter}
p=\frac{z_0^n-z_0}{1+z_0^{n+1}},
\qquad 1+z_0^{n+1}\neq0.
\end{equation}
The resulting parameter is real. Consequently, the ideal map preserves real-amplitude pure states and, more generally, the Bloch plane $v=0$.

\begin{figure}[pos=htbp]
\singlespacing
\centering
\includegraphics{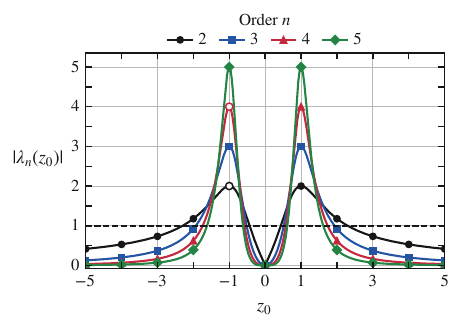}
\caption{Magnitude of the pure-state fixed-point multiplier for orders $n=2,3,4,5$. For each real $z_0$, the protocol parameter $p$ is chosen according to Eq.~\eqref{eq:reference_parameter}, and $\lambda_n(z_0)=f_p'(z_0)$ is evaluated with $p$ held fixed during differentiation. The dashed line at $\lvert\lambda_n\rvert=1$ separates attracting ($\lvert\lambda_n\rvert<1$) from repelling ($\lvert\lambda_n\rvert>1$) fixed points for pure-state evolution. Filled markers show calculated values where $p$ is finite. For even $n$, open circles mark $\lvert\lambda_n\rvert=n$, the value approached as $z_0$ gets closer to $-1$. For even $n$, the point $z_0=-1$ itself is excluded because no finite $p$ exists. For odd $n$, Eq.~\eqref{eq:reference_parameter} gives $p=0$ there.}
\label{fig:reference_multiplier}
\end{figure}

The fixed-point multiplier simplifies to
\begin{equation}\label{eq:reference_multiplier}
f_p'(z_0)
=\frac{n z_0^{\,n-1}(1+z_0^2)}{1+z_0^{2n}}.
\end{equation}
Its magnitude $\lambda=|f_p'(z_0)|$, denoted $\lvert\lambda_n\rvert$ in Fig.~\ref{fig:reference_multiplier}, determines local stability against pure-state perturbations. An attractive fixed point, $\lambda<1$, suppresses sufficiently small deviations, whereas a repelling point, $\lambda>1$, amplifies them. The reference state therefore controls a trade-off between sensitivity and acceptance, with the multiplier shown in Fig.~\ref{fig:reference_multiplier}.

For the quadratic protocol ($n=2$), Table~\ref{tab:reference_states} lists two reference configurations on opposite sides of the stability threshold $\lambda=1$.

\begin{table}[pos=htbp]
\centering
\begin{tabular}{@{}cccccc@{}}
\hline
$z_0$ & $p$ & $\lambda$ & $q_0$ & $P_{\mathrm{acc},5}^{\mathrm{ideal}}$ & Fixed point \\
\hline
$1.5$ & $6/35$ $(0.171)$ & $156/97$ $(1.608)$ & $97/169$ $(0.574)$ & $3.4\times10^{-8}$ & repelling \\
$3$   & $3/14$ $(0.214)$ & $30/41$ $(0.732)$  & $41/50$ $(0.820)$  & $2.1\times10^{-3}$ & attracting \\
\hline
\end{tabular}
\caption{Fixed-point reference configurations for the quadratic protocol ($n=2$). For each reference state $|\psi(z_0)\rangle$, the parameter $p$ follows from Eq.~\eqref{eq:reference_parameter}, the multiplier $\lambda=|f_p'(z_0)|$ from Eq.~\eqref{eq:reference_multiplier}, and the single-round acceptance $q_0=q(z_0)$ from Eq.~\eqref{eq:pure_input_success}. The column $P_{\mathrm{acc},5}^{\mathrm{ideal}}$ gives the ideal complete-tree acceptance, Eq.~\eqref{eq:tree_acceptance_fixed}, at the largest depth used in the experiments, $k=5$. Exact values are followed by decimal approximations in parentheses.}
\label{tab:reference_states}
\end{table}

The two configurations realize the trade-off between sensitivity and acceptance discussed above. At $z_0=1.5$, the fixed point is repelling: to first order, a pure-state deviation from the reference state grows by a factor $\lambda\approx1.61$ per round, so the output responds strongly to implementation errors. This sensitivity comes at the cost of a lower single-round acceptance, which compounds over the $2^k-1$ postselections of a complete tree and reduces the ideal acceptance at $k=5$ to the order of $10^{-8}$. At $z_0=3$, the fixed point is attracting and $q_0=0.82$, so the complete-tree acceptance decreases much more slowly with depth and deeper trees remain accessible to sampling. Small pure-state deviations, however, contract by a factor $\lambda\approx0.73$ per round, so this configuration is less sensitive to an individual error and probes mainly its accumulated effect.

In the ideal protocol, every successfully completed tree returns the reference state $|\psi(z_0)\rangle$, independently of its depth. Any departure from this state therefore indicates an implementation error, and detecting it does not require reconstructing the fractal boundary. Conversely, the enhanced sensitivity at a repelling fixed point is a local property of the map and does not imply that the experiment measures the fractal geometry directly. Results obtained with this construction should accordingly be reported together with the reference state, protocol order, parameter, and multiplier.

\subsection{Joint assessment of fidelity and acceptance}

Conditional fidelity alone can give an incomplete account of the implementation because postselection may discard erroneous runs. We therefore assess the accepted output together with the probability of accepting the complete tree. For each depth $k$, let $N_{\mathrm{raw},k}$ count all attempted trees and $N_{\mathrm{acc},k}$ those for which every required postselection succeeds. The measured acceptance probability is
\begin{equation}\label{eq:measured_tree_acceptance}
P_{\mathrm{acc},k}^{\mathrm{meas}}
=\frac{N_{\mathrm{acc},k}}{N_{\mathrm{raw},k}}.
\end{equation}
Every attempted tree contributes to the raw count, whether or not it is accepted, and failed branches are not rebuilt within a trial. For the accepted output state $\rho_k$, we define the conditional fidelity with the reference state,
\begin{equation}\label{eq:benchmark_fidelity}
\mathcal F_k=\langle\psi(z_0)|\rho_k|\psi(z_0)\rangle.
\end{equation}
It can be obtained from Pauli expectation values or, more directly, by rotating the reference state into the computational basis before the final measurement. In the latter case, $\mathcal F_k$ is the probability of the reference outcome among accepted trials.

The acceptance ratio
\begin{equation}\label{eq:benchmark_acceptance_ratio}
A_k=\frac{P_{\mathrm{acc},k}^{\mathrm{meas}}}
{P_{\mathrm{acc},k}^{\mathrm{ideal}}}
\end{equation}
can exceed one, and deviations in either direction indicate altered postselection statistics. We combine the two quantities into the score
\begin{equation}\label{eq:balanced_benchmark}
Q_k=
\begin{cases}
\mathcal F_k\min(A_k,A_k^{-1}),&A_k>0,\\
0,&A_k=0.
\end{cases}
\end{equation}
Its ideal value is one. The acceptance factor penalizes equal multiplicative deviations above and below the ideal probability equally, and setting $Q_k=0$ when no tree is accepted keeps the score defined even when the conditional fidelity cannot be estimated. Because $Q_k$ is a product of the two factors, a reduced score can be traced back to the fidelity or to the acceptance by examining $\mathcal F_k$ and $A_k$ separately.

\subsection{Demonstration on quantum hardware}

To demonstrate that the procedure can be carried out on available devices, we implemented the quadratic protocol for the two reference configurations of Table~\ref{tab:reference_states} on IQM Emerald and Rigetti Cepheus-1-108Q, accessed via Amazon Braket, for tree depths $k=1$--$5$. The purpose of these runs is to show that the required quantities can be obtained in practice, not to characterize or compare the devices.

The circuits prepare each qubit with $R_y(2\arctan z_0)$ and implement the real-parameter final gate as $R_y(-2\arctan p)$. All measurements are deferred until the end, without intermediate reset or qubit reuse. A final $R_y(-2\arctan z_0)$ on the surviving qubit maps the reference state to $|0\rangle$, so outcome $0$ of the surviving qubit tests the reference projector, while all other outcomes of the tree must be $0$ for acceptance. The measured reference probability includes final-readout errors and equals the intrinsic fidelity of Eq.~\eqref{eq:benchmark_fidelity} only for an ideal reference measurement. We use this measured probability as $\mathcal F_k$ both in the experimental score and in the noise model below.

Several complete trees are executed in parallel on disjoint blocks of qubits within each circuit, ranging from $27$ trees on $54$ qubits at $k=1$ to a single tree on $32$ qubits at $k=5$. Each depth was run with $10^5$ shots ($2\times10^5$ at $k=5$), giving between $2.7\times10^6$ raw tree trials at $k=1$ and $2\times10^5$ at $k=5$. Statistical uncertainties are estimated from $10^4$ bootstrap resamples of complete hardware shots, keeping trees measured in the same shot together, and are reported as $99.73\%$ percentile intervals, equivalent in coverage to $3\sigma$.

\begin{figure}[pos=htbp]
\singlespacing
\centering
\includegraphics[width=\linewidth]{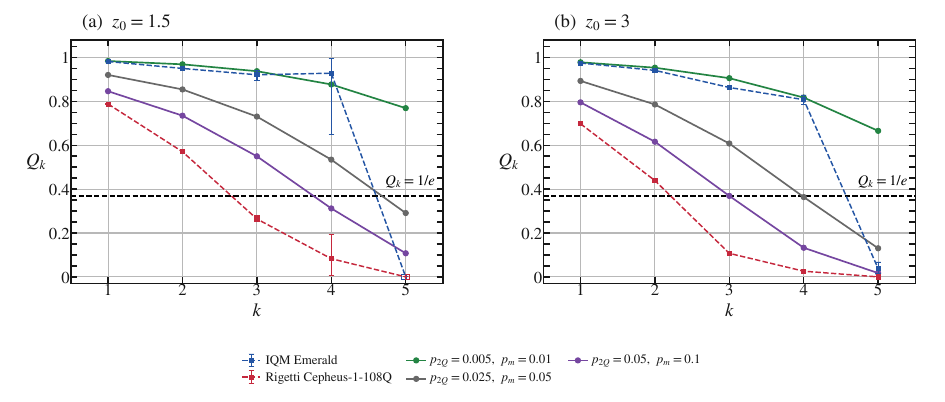}
\caption{Measured and simulated score $Q_k$ of the quadratic protocol as a function of tree depth $k$. Panels (a) and (b) use the repelling reference state $z_0=1.5$ and the attracting reference state $z_0=3$, respectively (Table~\ref{tab:reference_states}). Blue and red squares connected by dashed lines show results from IQM Emerald and Rigetti Cepheus-1-108Q, with $99.73\%$ percentile bootstrap intervals (equivalent to $3\sigma$). Solid lines with circles show exact infinite-shot predictions of the stochastic noise model of Eqs.~\eqref{eq:nl_two_qubit_noise} and \eqref{eq:nl_readout_noise} for the probabilities $(p_{2Q},p_m)$ given in the legend. Open markers at $k=5$ in panel (a) indicate that the ideal expected number of accepted trees within the available trials is below $10^{-2}$. These points reflect the finite shot budget rather than the device performance. The dashed horizontal line marks the illustrative threshold $Q_k=1/e$.}
\label{fig:nonlinear_benchmark_results}
\end{figure}

Figure~\ref{fig:nonlinear_benchmark_results} shows the resulting scores. For both reference states and both devices, $Q_k$ decreases with tree depth, as the number of gates and postselections grows. The deviations of $Q_k$ from its ideal value reflect changes in both the measured fidelity and the acceptance probability. For $z_0=3$, the accepted outputs maintain high fidelity, and the score reduction is mainly caused by acceptance falling below its ideal value. The sharp drop at $k=5$ is not a sampling effect, since about $4\times10^2$ accepted trees are expected ideally. It indicates a strong reduction of the measured acceptance on the $32$-qubit register, plausibly due to routing overhead and accumulated readout errors. For $z_0=1.5$, both contributions are visible, while the very low acceptance at greater depths limits the number of accepted trees and increases the statistical uncertainty. At $k=5$, the ideal acceptance for this configuration (Table~\ref{tab:reference_states}) corresponds to less than $10^{-2}$ expected accepted trees within the $2\times10^5$ trials, so the score at that depth carries little information. A larger shot budget would reduce the statistical uncertainty at large depths and allow $\mathcal F_k$ and $A_k$ to be estimated more precisely. For example, obtaining of order $10^2$ accepted trees at $k=5$ for $z_0=1.5$ would require about $3\times10^9$ raw trials under ideal acceptance. When no tree is accepted, the score is zero by definition. Preparation, gate, readout, and idling errors can affect both contributions, and these measurements alone do not distinguish their individual effects.

To illustrate how gate and measurement errors affect the score, we simulate the quadratic protocol using a local stochastic noise model. After each CNOT, we apply the two-qubit depolarizing channel
\begin{equation}
    \mathcal{E}_{2Q}(\rho) = (1-p_{2Q})\rho + \frac{p_{2Q}}{15} \sum_{P\in\{I,X,Y,Z\}^{\otimes2}\setminus\{I\otimes I\}} P\rho P,
    \label{eq:nl_two_qubit_noise}
\end{equation}
where $p_{2Q}$ is the total probability of a non-identity two-qubit Pauli error. Readout errors are modeled independently on each measured qubit by
\begin{equation}
    \mathcal{E}_{M}(\rho)=(1-p_m)\rho+p_mX\rho X,
    \label{eq:nl_readout_noise}
\end{equation}
which flips the computational-basis outcome with probability $p_m$. This channel acts immediately before the postselection measurements and before the final measurement, after the rotation that maps $|\psi(z_0)\rangle$ to $|0\rangle$. State preparation and single-qubit gates are ideal, and we assume unrestricted connectivity, so no routing operations are added. Since the noise acts locally on each step, the two subtrees entering a step are independent and identical, and the complete trees can be evaluated through an exact density-matrix recursion on two qubits. This gives the infinite-shot values of $\mathcal F_k$, $A_k$, and $Q_k$, including readout errors in both the fidelity and the acceptance. The three settings $(p_{2Q},p_m)=(0.005,0.01)$, $(0.025,0.05)$, and $(0.05,0.1)$ illustrate the response to increasing noise and are not fitted to the devices. Differences from the measurements may reflect routing overhead and error sources absent from this model, such as idle decoherence, coherent errors, and crosstalk.

To illustrate how the score could be condensed into a single figure of merit, we adopt the level $Q_k=1/e$ as an example threshold, corresponding to a decrease of the score by one $e$-fold from its ideal value. We define the threshold depth $k_{1/e}$ as the largest depth for which $Q_k$, including the lower end of its $99.73\%$ interval, remains above this level at that depth and at all smaller depths. For IQM Emerald, this gives $k_{1/e}=4$ for both reference states. For $z_0=1.5$, this value is a lower bound, since the score at $k=5$ is limited by the shot budget. For Rigetti Cepheus-1-108Q, it gives $k_{1/e}=2$ for both reference states. The choice of threshold is conventional, and a different level would generally give different depths. These values illustrate how the results can be summarized and are not intended to characterize the devices.

The construction extends directly to higher orders using the corresponding fixed-point parameter and ideal tree acceptance. Varying $n$ and $k$ then changes the width, gate count, and postselection overhead of the circuits, in the spirit of volumetric approaches to device testing~\cite{BlumeKohout2020Volumetric}. The resulting score reflects the combined effects of preparation, gates, measurement, and any storage required by the chosen schedule. Additional measurements or controlled comparisons are needed to attribute a deviation to a particular error source.

\section{Conclusions}

We investigated the fractal geometry of generalized iterated nonlinear quantum protocols and its response to preparation and gate noise. For orders $n=2$--$5$, moderate initial mixing preserves the main parameter-dependent trends of the boundary dimensions where the fractal structure remains resolved. The estimated dimension of the three-dimensional quasi-Julia boundary is approximately one unit larger than that of the corresponding pure-state Julia set over much of the studied parameter range. Its extent inside the Bloch ball provides a geometric interpretation of the purity-driven transition.

For the weak coherent and depolarizing errors studied, dimension differences are generally comparable to or smaller than the regression uncertainties, even when the convergence regions deform. The $ZZ$ dynamics are exactly conjugate to an ideal protocol at a rotated parameter, explaining their connection to the ideal dimension landscape. Their effects differ: coherent errors preserve pure-state evolution, whereas repeated depolarization shifts the attracting states into the Bloch-ball interior. Finally, we examined the implementation costs and proposed a fixed-point diagnostic that combines conditional fidelity with acceptance statistics to assess the noisy protocol. Measurements on IQM Emerald and Rigetti Cepheus-1-108Q illustrate this diagnostic for two quadratic reference states, together with illustrative stochastic-noise simulations.

\section*{Data and code availability}

The data supporting the findings of this study will be made
publicly available. The code will be available at
\url{https://gitlab.com/portikattila/fraqtal/}.

\section*{Acknowledgements}

This work was supported by the Horizon Europe programme HORIZON-CL4-2022-QUANTUM-01-SGA via the project 101113946 OpenSuperQPlus100. We also acknowledge support by the National Research, Development and Innovation Office within the Quantum Information National Laboratory of Hungary (Grant No. 2022-2.1.1-NL-2022-00004) and within the National Research, Development and Innovation Fund, through the AI4QT project Nr. 2020-1.2.3-EUREKA-2022-00029.

\bibliographystyle{elsarticle-num}
\bibliography{references}

\end{document}